\documentclass[]{rsos}

\begin{document}

\title{Characterizing Robustness of Strategies to Novelty in Zero-Sum Open Worlds}

\author{
Mayank Kejriwal$^{1}$, Shilpa Thomas$^{1}$ and Hongyu Li$^{1}$}

\address{$^{1}$Information Sciences Institute, University of Southern California, 4676 Admiralty Way, Ste. 1001, Marina del Rey, CA 90292, United States of America\\
}

\subject{Computer Science, Game Theory}


\corres{Mayank Kejriwal\\
\email{kejriwal@isi.edu}}

\begin{abstract}
In open-world environments, artificial agents must often contend with novel conditions that deviate from their training or design assumptions. This paper studies the robustness of fixed-strategy agents to such novelty within the setting of two-player zero-sum games. We present a general framework for characterizing the impact of environmental novelties, such as changes in payoff structure or action constraints, on agent performance in two distinct domains: Iterated Prisoner’s Dilemma (IPD) and heads-up Texas Hold’em Poker. Novelty is operationalized as a perturbation of the game's rules or scoring mechanics, while agent behavior remains fixed. To measure the effects, we introduce two metrics: \textit{per-agent robustness}, quantifying the relative performance shift of each strategy across novelties, and \textit{global impact}, summarizing the population-wide disruption caused by a novelty. Our experiments, comprising 30 IPD agents across 20 payoff matrix novelties and 10 Poker agents across 5 rule-based novelties, reveal systematic patterns in robustness and highlight certain novelties that induce severe destabilization. The results offer insights into agent generalizability under perturbation and provide a quantitative basis for designing safer and more resilient autonomous systems in adversarial and dynamic environments.
\end{abstract}


\maketitle

\section{Background and Related Work}\label{sec:background}

The recent maturation of autonomous systems has led to a growing recognition of \emph{open-world learning}, defined as the capacity of artificial agents to maintain or regain performance when novelties are introduced into their environment \cite{SAILON2021, boult2019learning, opp8}. While there exists a long-standing tradition of research focusing on adjacent problems, such as anomalous example detection \cite{main4}, change-point analysis \cite{killick2014changepoint}, and zero-shot or few-shot generalization \cite{ZS1, ZS2, main20}, the study of agents operating in complex, stochastic environments under rule-based or payoff-based perturbations remains significantly underdeveloped. Previous work has often focused on the detection of the "unknown unknown" \cite{opp11} or surveying steps toward open-world recognition \cite{boult2019learning, main5}. However, advanced open-world learning that evaluates the interplay between fixed strategic behavior and environmental mutation in zero-sum settings is still almost unheard of in the literature.

In this study, we do not attempt to implement new learning or adaptation algorithms. Instead, our goal is to provide rigorous empirical evidence for the \emph{necessity} of such learning by demonstrating the inherent fragility of fixed strategies. We show that agents display highly varied levels of robustness under different classes of novelty, emphasizing that a strategy successful in a closed-world benchmark may fail catastrophically when game-theoretic assumptions are violated \cite{main1, main2}. By characterizing how agents behave across a wide spectrum of injected novelties, ranging from payoff shifts in the Prisoner's Dilemma \cite{main9, eval7} to action constraints in Poker \cite{main11, main12}, we highlight a critical gap in current AI evaluation. Our work provides a domain-agnostic methodology \cite{eval5, eval8} to analyze robustness and global impact, serving as a foundational step toward the broader question of how agents must be modified to thrive in truly open worlds. This study-slash-methodology occupies a unique niche, as to our knowledge, a comparative cross-domain analysis of fixed-strategy robustness in response to systematic novelty has not been previously detailed in the academic record.

Artificial agents are increasingly expected to operate beyond tightly scoped benchmarks and into \emph{open worlds}---settings where unforeseen changes in rules, pay-offs or environment dynamics can arise without warning \cite{main5, main8}. Robust behaviour under such \emph{novelty} is particularly critical in \emph{competitive} arenas, where any brittleness may be exploited by an adversary \cite{main1, main14}.  

Zero-sum games furnish a principled, mathematically grounded lens for studying these issues because (i) payoffs have a clear conservation property, (ii) equilibrium concepts such as Nash optimality are well-defined, and (iii) agents’ objectives are perfectly opposed, making performance degradation easy to detect.  Yet, despite decades of work on strategic learning and opponent modelling \cite{main13}, comparatively little empirical evidence exists on how \emph{fixed} strategies respond when the underlying game itself mutates after deployment \cite{main2}.

The DARPA SAIL-ON programme has recently catalysed research on \emph{open-world learning}, formally separating the ability to \textit{detect}, \textit{characterise}, and \textit{adapt to} novelty \cite{SAILON2021, boult2019learning, opp8}.  Our focus in this article is on the second capability ({characterisation}) with specific emphasis on measuring how much performance of entrenched strategies degrades once a novelty is introduced \cite{eval9, opp7}.  Rather than endowing agents with explicit adaptation mechanisms, we treat each agent as immutable, echoing real-world situations in which updating a fielded system may be impossible or too slow to mitigate risk \cite{main15}.  This framing naturally raises two questions:

\begin{enumerate}
    \item[\textbf{Q1}] \emph{How can we quantify the robustness of a particular strategy to an arbitrary novelty?}
    \item[\textbf{Q2}] \emph{How disruptive is a given novelty when averaged over a diverse population of strategies?}
\end{enumerate}

To address these questions we study two canonical, but qualitatively distinct, zero-sum domains.  The first is the \emph{Iterated Prisoner’s Dilemma} (IPD), long used to probe cooperation and retaliation dynamics \cite{axelrod1980effective}.  IPD offers clean, interpretable pay-off matrices that can be perturbed directly, yielding twenty systematic novelties that range from benign affine transformations to complete reversals of the canonical temptation hierarchy \cite{main9, eval7}.  The second domain is \emph{heads-up Texas Hold-’em Poker}.  Unlike IPD, Poker is partially observable, features stochastic state transitions, and possesses a far richer action space.  We inject five novelties that reorder hand rankings, swap hole cards, or constrain betting actions, thereby stressing different facets of strategic reasoning \cite{main11, main12}.  By exploring both domains we demonstrate that our analysis framework is not artefact-specific but transfers across games with fundamentally different information structures.

Our methodology revolves around an \textbf{agent matrix} representation \cite{main4}.  For a fixed set of agents, we run round-robin tournaments both \emph{before} and \emph{after} novelty injection, recording each bilateral match-up as an element of a square matrix.  Element-wise comparisons yield (i) a per-agent robustness score, capturing the mean directional change in an agent’s win ratio or cash share; and (ii) a global novelty-impact statistic, capturing the magnitude and variability of performance shifts across the entire population \cite{eval6, eval4}.  We accompany these metrics with significance analyses and visualisations (e.g.\ heat-maps and boxplots) to surface macro-level phenomena such as shifts in dominance hierarchies or emergence of pathological equilibria.

\medskip
\noindent\textbf{Contributions.}  The main contributions of this work are four-fold:

\begin{enumerate}
    \item We introduce a \emph{domain-agnostic} matrix formalism and two complementary metrics—\textit{Per-Agent Robustness} and \textit{Global Impact}---for quantifying the effects of novelty in zero-sum competitive settings \cite{eval5, eval8}.
    \item We construct the largest to-date experimental corpus of novelty injections in IPD (30 agents $\times$ 20 novelties) and heads-up Poker (10 agents $\times$ 5 novelties), releasing code, parameter files, and result matrices for reproducibility.\footnote{Data and scripts for replicating the work are available at \url{https://github.com/anonymous/robust-novelty}.}
    \item Through comprehensive statistical tests we reveal (a) which canonical IPD agents remain surprisingly resilient, (b) which Poker strategies are susceptible to seemingly innocuous rule tweaks, and (c) how certain novelties flatten otherwise skewed competitive landscapes.
    \item We provide practical guidelines for researchers designing evaluation suites where robustness to environmental change is a first-class objective \cite{goss2023polycraft}, and we discuss implications for deploying real-world autonomous systems without closed-loop retraining.
\end{enumerate}

\noindent The remainder of this article is organised as follows.  Section~\ref{sec:novelty} formalises novelty in zero-sum open worlds \cite{main20, zhang2018overview} and details the two game domains, including defining our robustness and impact metrics.  Section~\ref{sec:materials_methods} describes the experimental setup, including agent libraries, novelty parameterisations, and statistical protocols.  Results for IPD and Poker are presented in Section~\ref{sec:results}.  Section~\ref{sec:discussion} discusses broader implications, limitations, and avenues for future work.  Finally, Section~\ref{sec:conclusion} summarises key findings and outlines next steps.

\section{Novelty in Zero-Sum Open Worlds}\label{sec:novelty}

To study novelty in open worlds, we devise experiments in two game-theoretic domains: \textit{iterated prisoner's dilemma (IPD)} and \emph{heads-up Poker (Texas hold 'em)}. 

{\bf Iterated Prisoner's Dilemma (IPD).} The Prisoner's Dilemma is a canonical $2 \times 2$ non-zero-sum game (often normalized to a zero-sum tournament frame in our analysis) where each player $i \in \{1, 2\}$ chooses an action $x_i \in \{C, D\}$, representing cooperation and defection respectively. The payoffs are defined by a matrix $M$, where $M(x_1, x_2) = (p_1, p_2)$ denotes the reward for each player. In the classic setting, the payoffs satisfy the hierarchy $T > R > P > S$, where $T$ is the temptation to defect, $R$ is the reward for mutual cooperation, $P$ is the punishment for mutual defection, and $S$ is the ``sucker's payoff'''. Formally, a single stage of the game is defined such that defection is a strictly dominant strategy, leading to the unique Nash equilibrium $(D, D)$, despite $(C, C)$ yielding a higher collective payoff. 

The iterated version (IPD) extends this to a sequence of $T$ rounds. At each round $t \in \{1, \dots, T\}$, players choose actions based on the history of previous play $H_{t-1} = \{(x_{1,1}, x_{2,1}), \dots, (x_{1,t-1}, x_{2,t-1})\}$. This repetition allows for the emergence of reciprocal strategies, as agents can condition their current behavior on the past actions of their opponent to punish defection or reward cooperation.

\begin{figure}
\centering
\includegraphics[width=6cm]{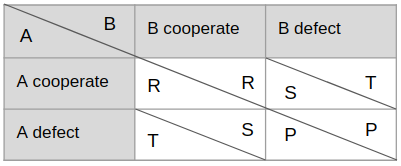}
\caption{Payoff matrix for the original Prisoner's Dilemma game, which serves as the baseline for our novelty injections.}
\label{fig:payoff}
\end{figure}

The IPD framework is particularly suited for novelty studies because the environmental dynamics are succinctly captured by the four payoff parameters. To test the effects of novelty, we admit out-of-distribution injections at the level of the payoff matrix. By perturbing the values or the ordering of $\{R, S, T, P\}$, we can transition the game from a coordination-friendly regime to one where the fundamental strategic incentives are inverted. Detailed descriptions of the core IPD agent strategies are provided in Appendix A. 

\textbf{Heads-Up Texas Hold’em Poker.} The second domain we consider is heads-up Texas Hold’em Poker, which introduces several complexities absent from IPD. While IPD is a game of perfect information regarding the game state (if not the opponent's intent), Poker is an \textit{imperfect information game}. The state space is defined by a deck of 52 cards, where each player $i$ is dealt two private ``hole" cards $c_i$, and up to five communal cards $C_{pub}$ are dealt across four betting rounds: pre-flop, flop, turn, and river. 

Formally, Poker introduces three core challenges for robust strategic reasoning:
\begin{enumerate}
    \item \textbf{Stochasticity and Hidden Information:} The transition between game states is governed by a stochastic deck, and players must reason over the distribution of an opponent's possible hole cards (their \textit{range}) rather than a known state.
    \item \textbf{Multi-stage Decision Space:} Unlike the single-choice nature of a PD round, a Poker hand involves sequential betting. In each round $k$, an agent chooses an action $a \in \{\text{fold, call, check, raise, all-in}\}$, affecting the size of the \textit{pot} $P$. The utility is determined by the final pot share at the end of the river or when one player folds.
    \item \textbf{Asymmetric Payoffs:} The rewards in Poker are not fixed by a matrix but are dynamically determined by player aggression and hand strength relative to the opponent.
\end{enumerate}

These features make Poker an ideal testbed for evaluating strategy robustness in the face of rich state spaces. We introduce targeted novelty injections e.g., reordering hand rankings or swapping hole cards, to observe how strategies calibrated for standard Poker logic collapse under mutated rules. We model agent behavior using a fixed suite of ten strategies ranging from simple rule-based agents to more structured policies that adjust aggression based on hand rank categories or betting history. Detailed descriptions of these strategies across different rounds are provided in Appendix B.
To simulate novelty in Poker, we introduce targeted perturbations that alter either the semantics of game rules or the decision constraints imposed on players. These include: reordering the standard ranking of hands (e.g., making a straight beat a flush), modifying card rank hierarchies (e.g., inverting the value of face cards), restricting action space during critical betting phase

\subsection{Metrics for Quantifying Impact}\label{sec:metrics}


Given a set of $n \geq 2$ agents $A=\{a_1,\ldots, a_n\}$, and a sequence of zero-sum games (called a \emph{tournament}) in which exactly two agents can participate at any given time, we can compute an agent vector $\vec{a_i}$ for an agent $a_i \in A$ by computing a \emph{tournament score} for $a_i$ when it plays against any agent $a_j \neq a_i$. The vector has dimensionality $n-1$, since there are $n-1$ agents that $a_i$ can play against. An systematic way to represent all the agent vectors is to construct an agent matrix $\mathbf{A}=A \times A$, with a `dummy' value such as 1.0 on the diagonal. A cell $(a_i,a_j)$ is the tournament score achieved by $a_i$ when it plays in a tournament against $a_j$. Per this construction, the row corresponding to $a_i$ is referred to as $a_i$'s agent vector. 

The manner in which the tournament score is computed will depend on several factors, including the domain. One simple (and universally applicable) choice is the \emph{win ratio}, defined as the proportion of games in the tournament that an agent has won. The win ratio gives the tournament the same zero-sum flavor that the underlying game has. A higher win ratio for one player will always reduce the win ratio for the other player, since (not including ties), the two win ratios must always add to 1. More generally, while $\mathbf{A}$ is generally asymmetric, there is a special relation between the cells $(a_i,a_j)$ and $(a_j,a_i)$. Assuming a \emph{normalized} zero-sum metric such as win ratio (with another Poker-specific metric described below), the values in both cells must necessarily add to 1.0. This also justifies why we chose 1.0 as our diagonal dummy value, as it allows us to maintain this property.

Alternatively, in most real-world games such as Poker, other zero-sum metrics can also be defined. For example, we may choose not to compute the number of hands won by a player, but to only consider the total amount of money the player possesses at the end of the tournament, which is the true measure of decision-making success in such high-variance games. If each player starts with the same amount of money at the beginning of the tournament (the `buy in', to use Poker terminology), and assuming money never leaves the table, the total amount of cash, summed over all players, will always be a constant at any point in the tournament. However, a superior player will have ended up increasing their \emph{share} by the end of the tournament, which by definition, must come at the expense of one or more players. In \emph{heads-up Poker} (with only two players), it must come at the expense of the only other player.  The tournament score in such a setup would then be the ratio of the agent's final cash balance to the total cash. This is also a zero-sum metric, and must add to 1.0 for both players.

While we do not need to settle upon a single canonical way of defining a tournament score, even for a particular domain or world, we note that it could influence the interpretation of the results. Hence, the choice of tournament score, just like the choice of novelties and agent-set, are important empirical concerns. Nevertheless, in domains where each game's score is computed \emph{independently}, the win ratio is a sound choice for the tournament score. That is, unlike games like Poker where the cash balance carries forward, the score `resets' at the beginning of the next game in the tournament, and in theory, each of the two agents has a chance to win the game. The IPD domain is a good example. Note that agent \emph{performance} on each game in the tournament is not necessarily independent, since agents that take any `history' of play into account (as do most well-performing agents, including Tit-for-Tat) will learn something from the previous games, especially when playing against the same opponent in a repeated setting. 


With this formalism in place, we note that, in order to characterize agent robustness and novelty impacts, we will need to construct a separate `default' agent matrix $\mathbf{A}^-$ for the tournaments that are played using the default version of the world (without novelty), and another matrix $\mathbf{A}^+$ representing agent performance on tournaments with the novelty. Since there might be multiple novelties, the `with-novelty' matrix $\mathbf{A}^+$ will depend on the novelty (but the default matrix $\mathbf{A}^-$ will not). We illustrate the $30 \times 30$ pre-novelty matrix $\mathbf{A}^-$ for the 30 agents we used in this study (described further in \emph{Materials and Methods} and also in \emph{Appendix A}) in Figure \ref{fig:prenov}. 

\begin{figure}
\centering
\includegraphics[trim = 300 150 30 50, width=14cm,height=9cm]{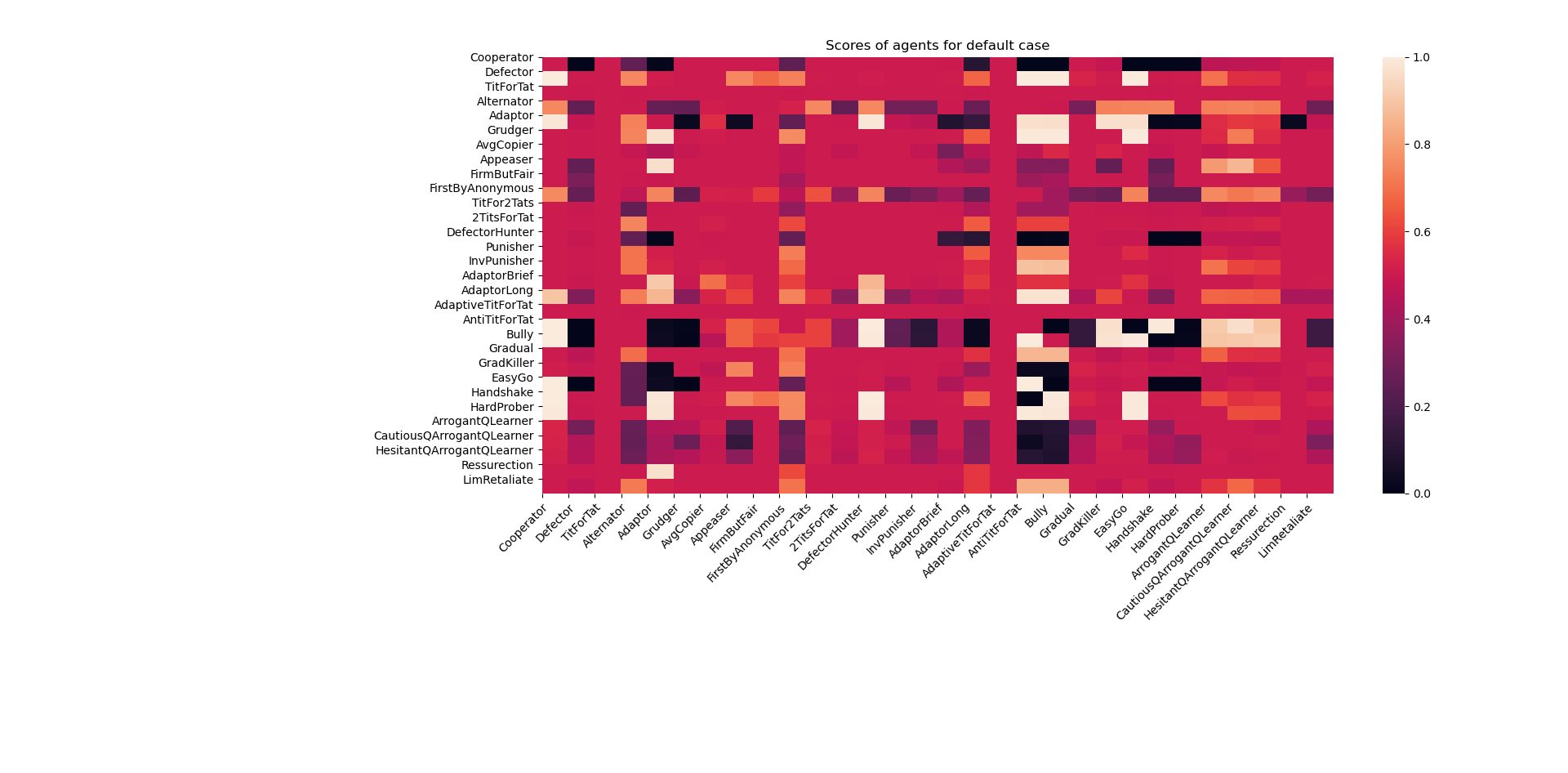}
\caption{Pre-novelty $30 \times 30$ agent matrix for IPD}
\label{fig:prenov}
\end{figure}

We assume a set of $k$ novelties $\mathbf{N}=\{N_1,\ldots,N_k\}$ and use a subscript when we are distinguishing a with-novelty agent matrix by novelty. For example, $\mathbf{A}_i^+$ represents the with-novelty matrix constructed for tournaments where the novelty $n_i$ was introduced. As detailed in \emph{Materials and Methods}, we devised $k=20$ novelties for the experiments in this paper. 

We present two metrics below, one of which (`per-agent robustness') is calculated across all novelties for each \emph{agent}, while the other (`global impact') is calculated across all agents for each \emph{novelty}. We assume that $\mathbf{A}$ and  $\mathbf{N}$ have been set by the experimenter, although knowledge of the novelties is withheld from the agents.

\subsection{Per-Agent Robustness}

The first metric we propose is \emph{Per-Agent Robustness}, which quantifies how resilient a given agent’s performance is to environmental novelty. Formally, given a set of $k$ novelties $\mathbf{N} = \{N_1, \ldots, N_k\}$, and a fixed set of agents $A = \{a_1, \ldots, a_n\}$, we compute for each agent $a_i \in A$ a robustness vector $\vec{r}_{a_i} \in \mathbb{R}^k$. Each element $r_{a_i}^{(j)}$ of this vector captures the mean tournament score of agent $a_i$ under novelty $N_j$, computed by averaging $a_i$'s normalized performance across all pairwise matchups against agents $a_j \ne a_i$ under that novelty.

To compute these scores, we rely on the agent matrix formalism described earlier. For a given novelty $N_j$, we first construct a tournament matrix $\mathbf{A}_j^+$ whose $(i, j)$-th entry records the tournament score of agent $a_i$ when playing against $a_j$. We then extract the $i$-th row of this matrix (excluding the diagonal), which represents $a_i$'s relative performance vector against the other agents under that novelty. The robustness score $r_{a_i}^{(j)}$ is computed as the mean of these entries:
\[
r_{a_i}^{(j)} = \frac{1}{n-1} \sum_{\substack{a_m \in A\\ m \ne i}} \mathbf{A}_j^+(i, m).
\]
This process yields a $k$-dimensional robustness vector for each agent. These vectors allow us to compare how consistently each agent performs across all injected novelties. An agent whose robustness vector exhibits low variance and values close to its pre-novelty performance can be considered robust to novelty. Conversely, agents whose robustness vectors show large declines or variability are more sensitive to environmental perturbations.

To summarize these trends visually, we construct a boxplot for each agent, aggregating their $k$ robustness values --- across all 30 agents and 20 novelties in the IPD domain. Robust agents are expected to have narrow boxplots centered around high scores, while vulnerable ones should show wide distributions or consistent drops. Hence, these plots reveal which agents are consistently strong performers and which are brittle in the face of changing rules. We further complement this analysis with statistical tests (detailed in Section~\ref{sec:materials_methods}) to assess whether the observed changes are statistically significant.




\subsection{Global Impact}

While per-agent robustness focuses on individual agent responses to novelty, the \emph{Global Impact} metric captures the overall disruptiveness of a novelty across an entire population of strategies. This allows us to rank novelties by their systemic effects, identifying which ones induce the largest shifts in competitive dynamics.

For a given novelty $N_j$, we construct two agent matrices: the baseline matrix $\mathbf{A}^-$ (pre-novelty) and the novelty-specific matrix $\mathbf{A}_j^+$ (post-novelty). Both matrices are of size $n \times n$, where $n$ is the number of agents. Because the diagonal entries are placeholders (set to 1.0), we focus only on the off-diagonal entries. Let $\Delta_j$ denote the set of absolute differences in corresponding entries between these two matrices:
\[
\Delta_j = \left\{ \left| \mathbf{A}_j^+(i, m) - \mathbf{A}^-(i, m) \right| \; \middle| \; i \ne m \right\}.
\]
The global impact of novelty $N_j$ is then defined as the mean of these differences:
\[
\text{Impact}(N_j) = \frac{1}{n(n-1)} \sum_{i \ne m} \left| \mathbf{A}_j^+(i, m) - \mathbf{A}^-(i, m) \right|.
\]
This scalar value captures the average perturbation induced by novelty $N_j$ across all agent match-ups. A higher global impact score implies that the novelty substantially alters relative agent performance, whereas a lower score suggests the novelty has little systemic effect.


Together with the per-agent robustness results, the global impact metric provides a dual perspective on the effects of novelty — one agent-centric, the other system-centric — enabling a richer characterization of robustness in adversarial open-world environments.




\section{Materials and Methods}\label{sec:materials_methods}

To evaluate the robustness of strategies in zero-sum open worlds, we implemented a simulation-based experimental framework across our two target domains. In both domains, we adhere to a methodology of fixed-strategy evaluation under environmental perturbation.

\begin{table}[h]
  \begin{center}
    \begin{tabular}{|p{1.0cm}|p{0.8cm}|p{0.8cm}|p{0.8cm}|p{0.8cm}|p{5.5cm}|} 
    \hline
  {\bf Novelty ID} & {\bf R'} & {\bf S'} & {\bf T'} & {\bf P'} & {\bf Notes} \\ 
  \hline
  {\it Default} & {\it R=6} & {\it S=0} & {\it T=10} & {\it P=1} & \\
  \hline \hline
  1 & R*100 & S*100 & T*100 & P*100 & All values multiplied by 100 \\ \hline
  2 & R*10 & S*10 & T*10 & P*10 & All values multiplied by 10 \\ \hline
  3 & R*10 & 0 & T*100 & 6 & Inequality still remains ($T>R>P>S$) but values have been randomly changed \\ \hline
  4 & 1 & 0 & 10 & 6 & $P>R$ and the inequality now becomes $T>P>R>S$ \\ \hline
  5 & 1 & 10 & 0 & 6 & Inequality is reversed to $T<R<P<S$ \\ \hline
  6 & 6 & 10 & 0 & 1 & $T<P<R<S$ \\ \hline
  7 & 6000 & 20 & 1 & 10000 & $T<S<R<P$ \\ \hline
  8 & 60 & 20000 & 1 & 100 & $T<R<P<S$ \\ \hline
  9 & 60 & 20000 & 20000 & 60 & $R=P<T=S$ \\  \hline
  10 & 60000 & 20 & 20 & 60000 & $R=P$, $T=S$ and $P>>T$ \\  \hline
  11 & 10 & 1 & 1 & 10 & $R=P>T=S$ \\ \hline
  12 & 100 & 100 & 100 & 100 & $R=P=T=S$ \\ \hline
  13 & 10 & 10 & 1000 & 1000 & $R=S$ and $T=P$ \\ \hline
  14 & 10 & 1000 & 10 & 1000 & $R=T$ and $S=P$ \\ \hline
  15 & 1000 & 1000 & 10 & 1000 & $R=S=P>T$ \\ \hline
  16 & 1000 & 10 & 1000 & 1000 & $R=T=P>S$ \\ \hline
  17 & 10 & 1000 & 1000 & 1000 & $S=P=T>R$ \\ \hline
  18 & 1000 & 1000 & 1000 & 10 & $R=S=T>P$ \\ \hline
  19 & 8520 & 0 & 1011 & 102 & $R>T>P>S$ \\ \hline
  20 & 6 & 15200 & 1110 & 1 & $S>T>R>P$ \\
  \hline
  \end{tabular}
    \medskip
    \caption{A summary of novelties considered in the IPD experimental study.}\label{tab:IPDnov}
    \end{center}
\end{table}

\begin{table}[h]
  \centering
  \small
  \begin{tabular}{|l|p{10cm}|}
    \hline
    \textbf{Agent} & \textbf{Strategic Summary} \\ \hline
    Raise & Doubles previous bet if cash allows; aggressive bias. \\ \hline
    Call & Matches any bet; passive persistence. \\ \hline
    Random & Stochastic action selection (fold, call, raise, all-in). \\ \hline
    Call-AT / Raise-AT & Conditioned on hole card strength (7-7+ or 10-A combinations). \\ \hline
    Raise-Aggressive & Multiplies previous bet by 10 for strong hole cards. \\ \hline
    Call-TopPair & Restricted to high-face hole cards (Jack to Ace). \\ \hline
    Call-Flop & Post-flop specialist; calls only on pairs or suit matches. \\ \hline
    Background V1/V2 & Categorical risk management; multi-stage policies for turn and river. \\ \hline
  \end{tabular}
  \caption{Summary of key strategic elements of the implemented Poker agents.}
  \label{tab:poker_agents}
\end{table}

\subsection{Iterated Prisoner's Dilemma (IPD) Setup}

In the IPD domain, we utilized the Axelrod Python library, a recognized benchmark for reproducible game-theoretic research (\url{https://axelrod.readthedocs.io/}) \cite{axelrod1980effective}. This repository consists of a library of over 200 agents with different strategies based on which the agent decides to either cooperate or defect. We cloned and modified this repository to allow for injection of novelties, and to facilitate calculation of average tournament scores for agents when they play against each other. 

We selected a diverse cohort of 30 agents representing a wide spectrum of strategic logic, including classic tit-for-tat variants, stochastic players, and memory-intensive retaliatory strategies \cite{press2012iterated, li2011engineering}. Technical descriptions for each of these agents can be found in Appendix A. For each agent pair, we conducted tournaments consisting of 100 rounds of the game. Across 20 simulated novelty scenarios (Table~\ref{tab:IPDnov}), we systematically modified the payoff matrix $\{R, S, T, P\}$ from the baseline configuration (Figure~\ref{fig:payoff}) \cite{mittaloptimal}. These modifications succinctly describe how the game environment transforms, helping us understand the effects of novelty on player performance and the impact of their respective strategies on the score. An systematic way to represent all the agent vectors is to construct an agent matrix $\mathbf{A}=A \times A$, as illustrated in Figure \ref{fig:prenov}. 

\subsection{Heads-Up Poker Setup}

In the Poker domain, we developed a custom heads-up Texas Hold'em environment. Our agent library for Poker consists of 10 distinct strategies, described in Table~\ref{tab:poker_agents}. These include rule-based heuristic agents (e.g., Raise, Call), threshold-driven aggression models (e.g., Call-AT, Raise-Aggressive), and more complex baseline players (Background V1 and V2) that maintain multi-stage betting policies across pre-flop, flop, turn, and river rounds. Specifically, according to each phase (pre-flop, flop, turn, and river), agents have different strategies. Only background agents V1 and V2 have strategies in turn and river rounds; the other agents, in both turn and river rounds, use the same strategies as in the flop round. Detailed descriptions of these strategies across different rounds are provided in Appendix B. Summary descriptions are provided in Tables~\ref{tab:poker_pre_strat} and~\ref{tab:poker_flop_strat}.

\begin{table}[h]
\centering
\small
\begin{tabular}{|l|l|p{8cm}|}
\hline 
\textbf{ID} & \textbf{Agent} & \textbf{Pre-Flop Round Description} \\ \hline 
 1 & Raise & If agent has amount of cash larger than twice of the previous bet, it would raise this amount. \\  
 2 & Call & If agent has amount of cash larger or equal to the previous bet on the table, it calls that amount. \\
 3 & Random & The agent takes random actions if those actions are allowable during the current phase. \\
 4 & Call-AT & The agent would call only if it gets a pair between 7 and Ace, or 10-Ace combinations. \\
 5 & Raise-AT & The agent would double the previous bet only if it gets high-value hole cards. \\
 6 & Raise-Aggressive & The agent raises the bet aggressively (10x) for strong hole cards. \\
 7 & Call-TopPair & The agent calls only for hole cards between Jack and Ace. \\
 8 & Call-Flop & The agent calls only for pairs between 7 and Ace, or 10-Ace combinations. \\
 9 & Background V1 & Categorical percent limits for betting based on hole card strength. \\
 10 & Background V2 & Strategy for pre-flop round is same as agent v1. \\
 \hline
\end{tabular}
\caption{Poker Pre-Flop Round Strategies}
\label{tab:poker_pre_strat}
\end{table}

\begin{table}[h]
\centering
\small
\begin{tabular}{|l|l|p{8cm}|}
\hline 
\textbf{ID} & \textbf{Agent} & \textbf{Flop Round Description} \\ \hline 
 1 & Raise & Doubles the bet on the table if cash allows. Folds if opponent goes all-in.  \\  
 2 & Call & Calls with the bet on the table. Folds if opponent goes all-in. \\
 3 & Random & Takes random allowable actions. \\
 4 & Call-AT & Calls with the bet on the table. Folds if opponent goes all-in. \\
 5 & Raise-AT & Doubles the bet on the table. Folds if opponent goes all-in. \\
 6 & Raise-Aggressive & Raises the bet 10 times. Folds if opponent goes all-in. \\
 7 & Call-TopPair & Calls with the bet on the table. Folds if opponent goes all-in. \\
 8 & Call-Flop & Calls only if current hands have $\ge 3$ cards with same values or same suits. \\
 9 & Background V1 & All-in if in top 1-3 pairs. Strategic folding under specific thresholds. \\
 10 & Background V2 & Strategy for flop round is same as agent v1. \\
 \hline
\end{tabular}
\caption{Poker Flop Round Strategies}
\label{tab:poker_flop_strat}
\end{table}

For each novelty injection, we simulated 1,000 hands per agent pairing to ensure statistical stability in the face of card-draw stochasticity. We implemented five rule-based novelties, detailed further in Appendix B and enumerated briefly below:
\begin{itemize}
    \item \textbf{Exchange Hand}: Hands are moved between active players before showdown.
    \item \textbf{Reorder Hand Ranking}: Shuffles the hierarchy of hands (e.g., High Card beats Full House).
    \item \textbf{Reorder Number Ranking}: Shuffles individual card values (e.g., 5 beats Ace).
    \item \textbf{Royal Texas}: Limits the deck to cards with values 10 and above.
    \item \textbf{Action Constraints}: Restricts actions to Call/Fold (pre-flop) or All-in/Fold (flop).
\end{itemize}
These novelties are injected into the Poker simulator to characterize the robustness of strategies under out-of-distribution rule mutations.

\subsection{Significance Analysis Protocol}

To quantify the significance of the observed shifts in performance, we employed a one-sample t-test for each agent across the injected novelties. The null hypothesis assumes that the mean per-agent robustness score under novelty is statistically indistinguishable from the baseline performance. We evaluated these tests at 95\% and 99\% confidence levels to identify strategies with significant deviations from their expected performance. All simulation code and result matrices were processed using standard scientific Python stacks (NumPy, SciPy, and Matplotlib).

\newpage

\section{Results}\label{sec:results}

\subsection{Iterated Prisoner's Dilemma}

\begin{figure}
\centering
\includegraphics[trim = 100 0 100 100, width=13cm,height=9cm]{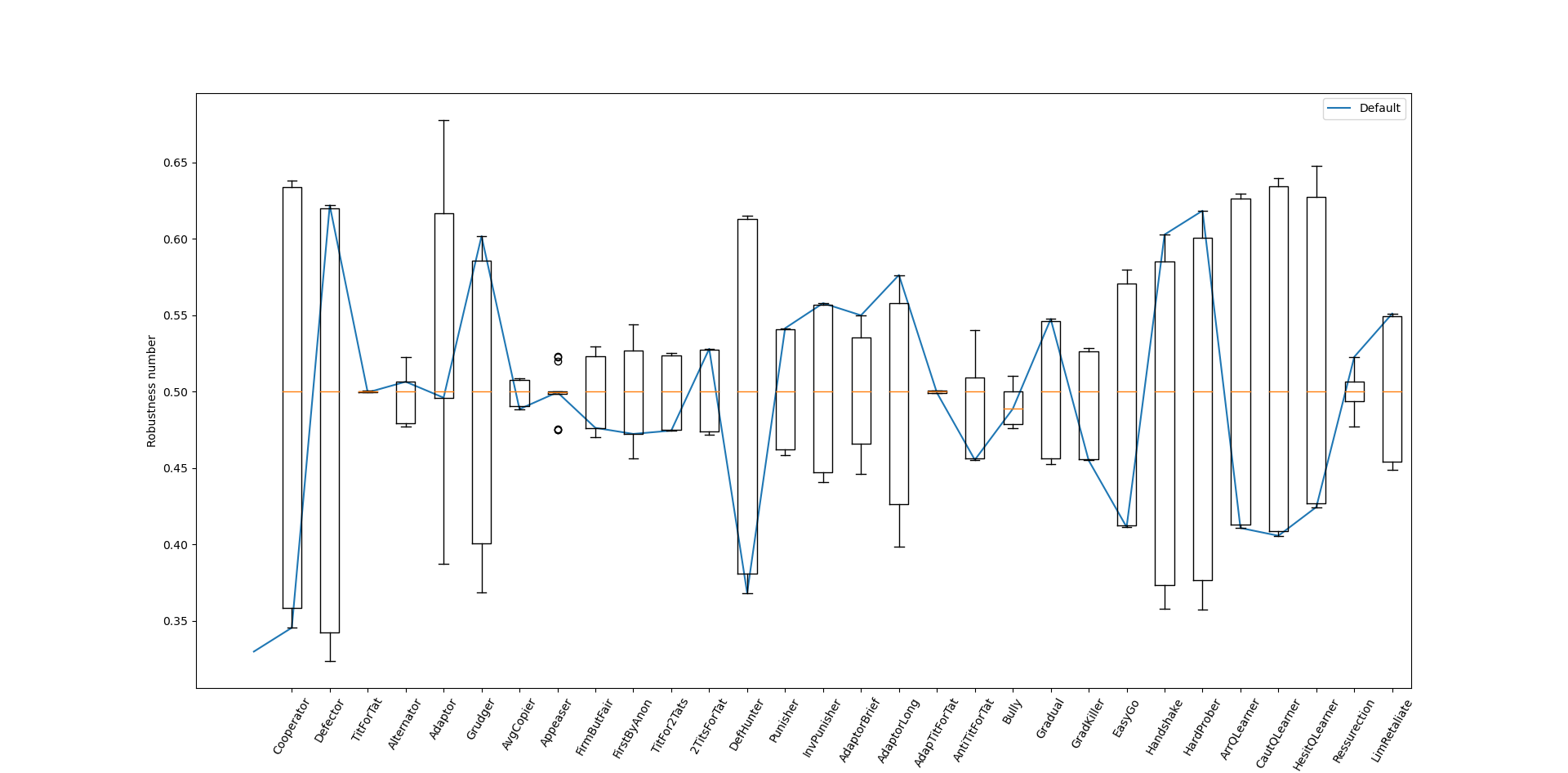}
\caption{Distribution of per-agent robustness values for 30 agents. Each box plot summarizes the distribution of robustness metrics calculated across 20 distinct novelty scenarios.}
\label{fig:agent}
\end{figure}

Figure \ref{fig:agent} presents the distribution of per-agent robustness across the 30 evaluated agents, where each box plot represents the performance variance of a specific agent across 20 distinct novelty scenarios. We observe substantial heterogeneity in agent performance, indicating that an agent's architectural design and strategy significantly dictate its capacity to maintain performance in unforeseen environments. This variability is most pronounced in baseline strategies such as the pure Cooperator and Defector, which exhibit the highest variance, with robustness values spanning a wide range from approximately 0.35 to 0.65. This suggests that these simplistic strategies are highly sensitive to the specific nature of the environmental novelty.

Conversely, more sophisticated strategies like Tit-for-Tat and its variants demonstrate remarkably narrow distributions, suggesting a high degree of consistency and reliable robustness across diverse novelty conditions. Other agents, such as Punisher, Inverse Punisher, and Adapter Brief, occupy a middle ground; while they exhibit greater variance than Tit-for-Tat, their robustness remains considerably more stable than that of the pure Cooperator or Defector. Collectively, these results underscore a central finding of this study: the choice of agent strategy is a critical determinant of system robustness. The ability to navigate novelty that was not explicitly anticipated during the design phase is not a uniform trait but is instead strongly modulated by the agent's underlying strategic logic. A pragmatic interpretation or application of this finding is that different agents will require different degrees of \textit{adaptation} to novelties, at least on average, and this should be taken into consideration when desgining a novelty adaptation system. 

In terms of statistical significance, analysis of the per-agent robustness values across the 20 novelty scenarios reveals that the majority of the population was affected by the environmental shifts. Out of the 30 agents evaluated, 27 exhibited significant changes in robustness at the 99\% confidence level, while one agent showed significance at the 95\% confidence level. In contrast, the Appeaser and AdapTitForTat agents did not exhibit statistically significant shifts in their post-novelty robustness, indicating a unique level of invariance to the introduced novelty scenarios.

\begin{figure}
\centering
\includegraphics[trim = 300 150 30 50, width=14cm,height=9cm]{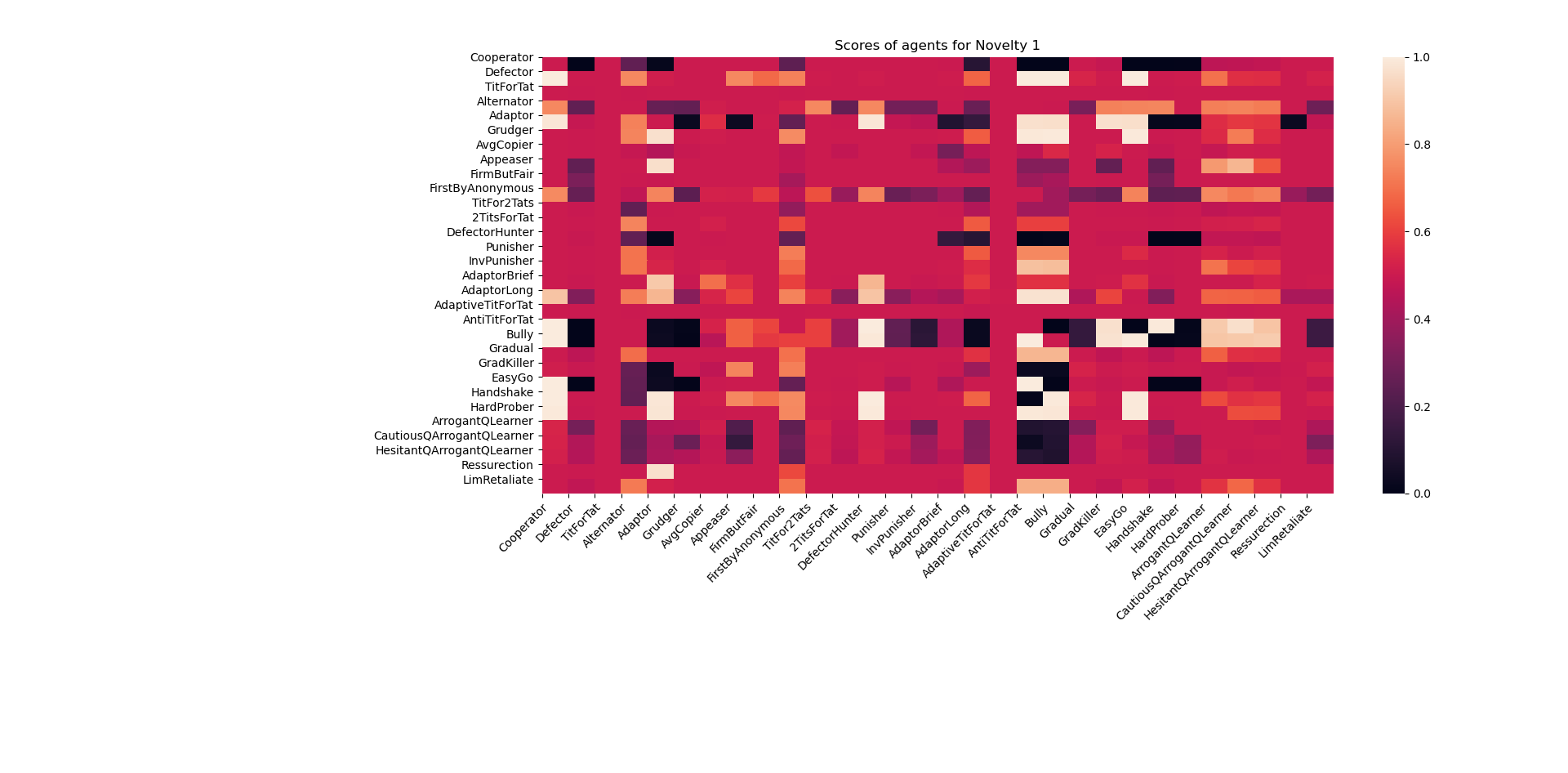}
\caption{Post-novelty payoff matrix under Novelty 1 (Linear Scaling), demonstrating invariant agent performance compared to the baseline.}
\label{fig:board1}
\end{figure}

\begin{figure}
\centering
\includegraphics[trim = 300 150 30 50, width=14cm,height=9cm]{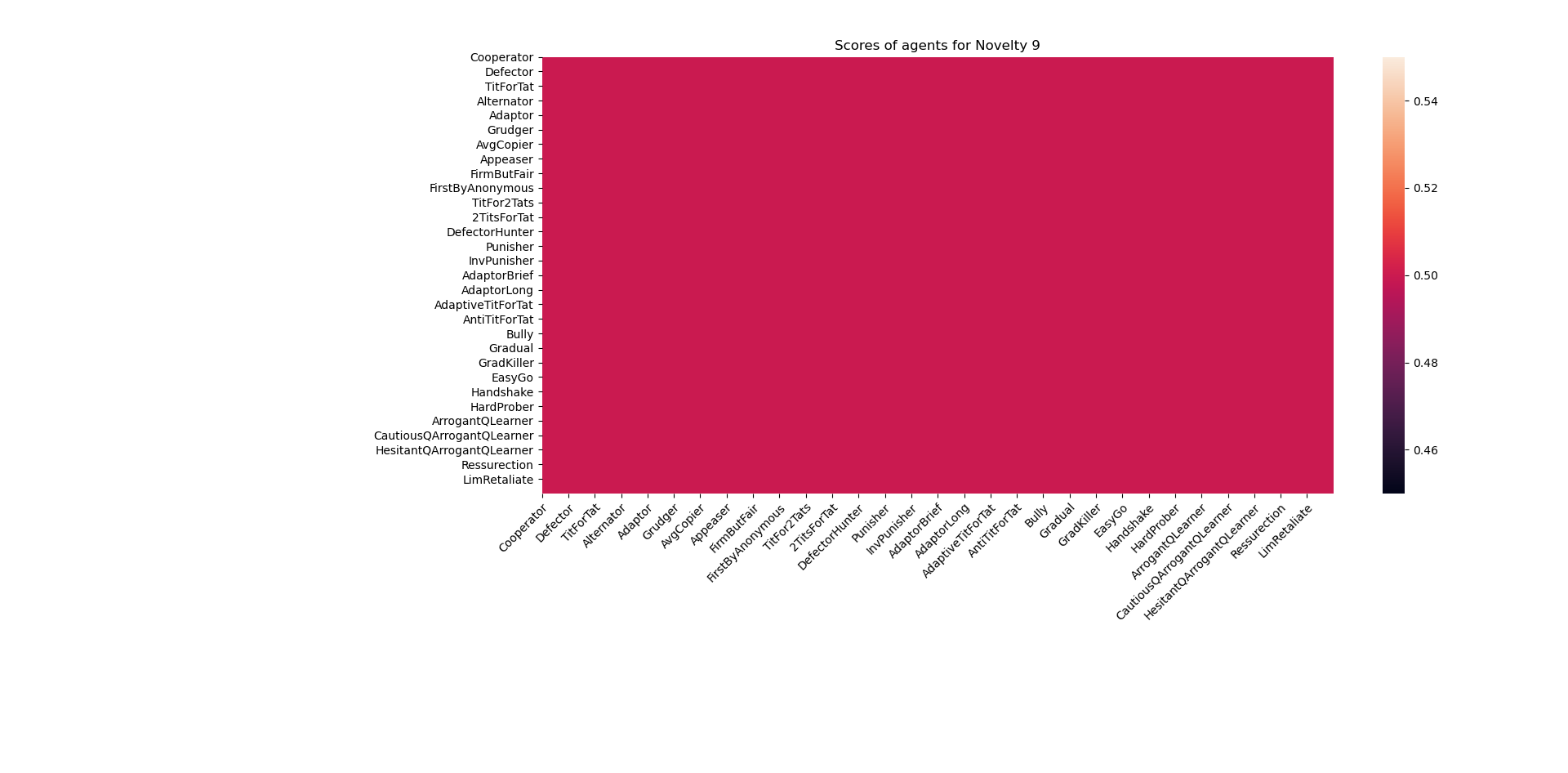}
\caption{Post-novelty payoff matrix under Novelty 9 ($R=P, S=T$), where strategy convergence leads to uniform payoffs across the agent population.}
\label{fig:board2}
\end{figure}

\begin{figure}
\centering
\includegraphics[trim = 300 150 30 50, width=14cm,height=9cm]{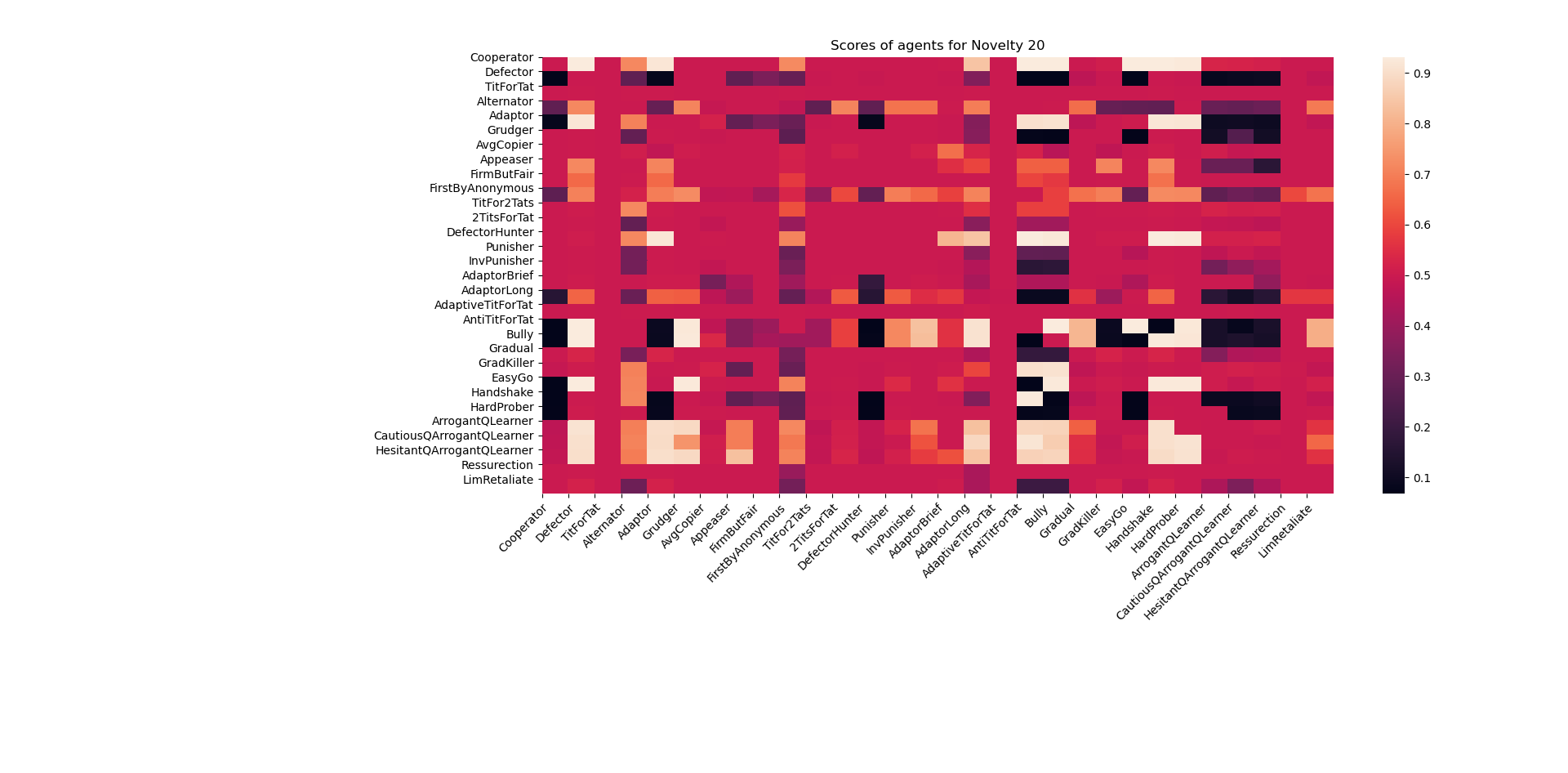}
\caption{Post-novelty payoff matrix under Novelty 20 ($S>T$), illustrating the performance advantage of cooperative strategies in a modified incentive landscape.}
\label{fig:board3}
\end{figure}

The impact of specific environmental modifications is further illustrated through the agent interaction matrices in Figures \ref{fig:board1}, \ref{fig:board2}, and \ref{fig:board3}.

Novelty 1 represents a linear scaling of the default pre-novelty payoff matrix, where each value is multiplied by a factor of 100. Since this transformation preserves the relative incentives and the Nash equilibrium of the underlying game, it is expected to have no functional effect on agent behavior. Indeed, the average scores in the post-novelty phase remain identical to those in the pre-novelty baseline, as shown in the interaction matrix in Figure \ref{fig:board1}.

In Novelty 9, the payoff structure is modified such that $R=P$ and $S=T$. Under these conditions, the rewards for cooperation and defection are equalized, rendering the choice of strategy irrelevant to the resulting payoff. Consequently, all agents receive an identical average score of 0.5 regardless of their strategic orientation, a result clearly visible in the uniform distribution shown in Figure \ref{fig:board2}.

Novelty 20 introduces a fundamental shift in the game's incentive structure by setting $S>T$ while maintaining $R>P$. By reversing the traditional inequality between the "sucker's payoff" ($S$) and the "temptation to defect" ($T$), cooperation is transformed into the dominant strategy. As shown in Figure \ref{fig:board3}, agents that exhibit a higher propensity for cooperation achieve significantly higher scores in this environment compared to those favoring defection.

\begin{figure}
\centering
\includegraphics[trim = 100 0 100 100, width=13cm,height=9cm]{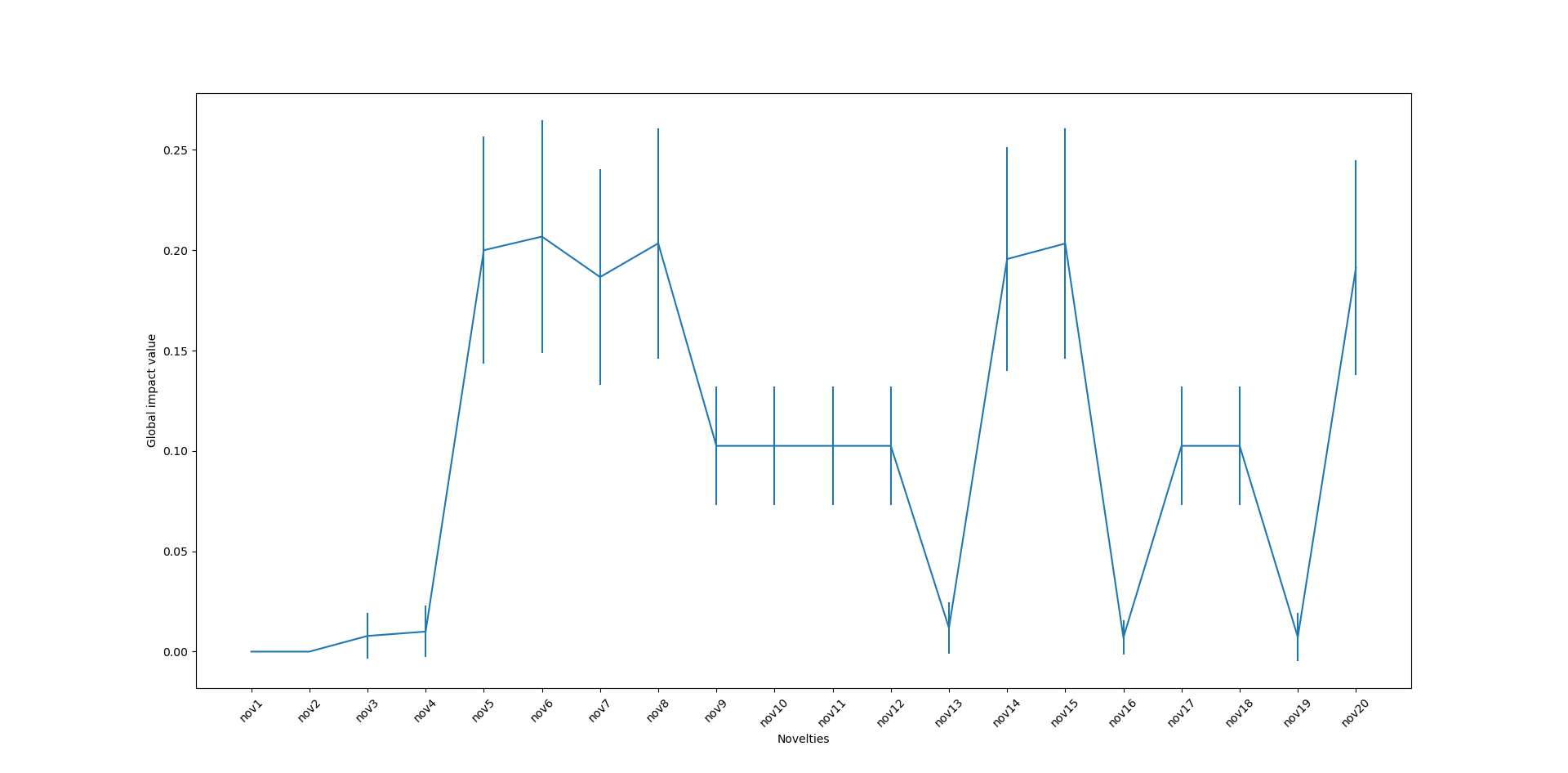}
\caption{The global impact of each novelty in the IPD domain (with standard errors).}
\label{fig:impact}
\end{figure}

Figure \ref{fig:impact} presents the global impact of each novelty, aggregated across the entire agent population. In contrast to the per-agent robustness distribution shown in Figure \ref{fig:agent}, which highlights how different agents respond to environmental shifts, this figure illustrates how the specific nature of the novelty itself dictates the magnitude of the disruption. Here, the x-axis represents each of the 20 distinct novelty scenarios, while the y-axis indicates the resulting global impact value.

We observe that the choice of novelty is a critical factor in determining the overall impact on the system. For instance, novelties 1, 2, and 3 exhibit minimal impact, suggesting that these specific environmental modifications do not fundamentally alter the strategic landscape or agent performance. Conversely, novelties 5-8, 14, 15, and 20 result in high global impact, representing scenarios where the agents' pre-existing strategies are significantly challenged. Other modifications, such as novelties 10, 11, 17, and 18, result in moderate impact levels. While the standard errors are discernible, they are notably narrower than the wide variance observed in the per-agent robustness boxes. These findings parallel our observations regarding agent strategy: just as the choice of agent significantly influences robustness, the specific characteristics of the novelty itself are a primary determinant of whether a global impact is observed across the population.

\subsection{Texas Hold'em Heads-Up Poker}

\begin{figure}[h]
\centering
\includegraphics[scale=0.8]{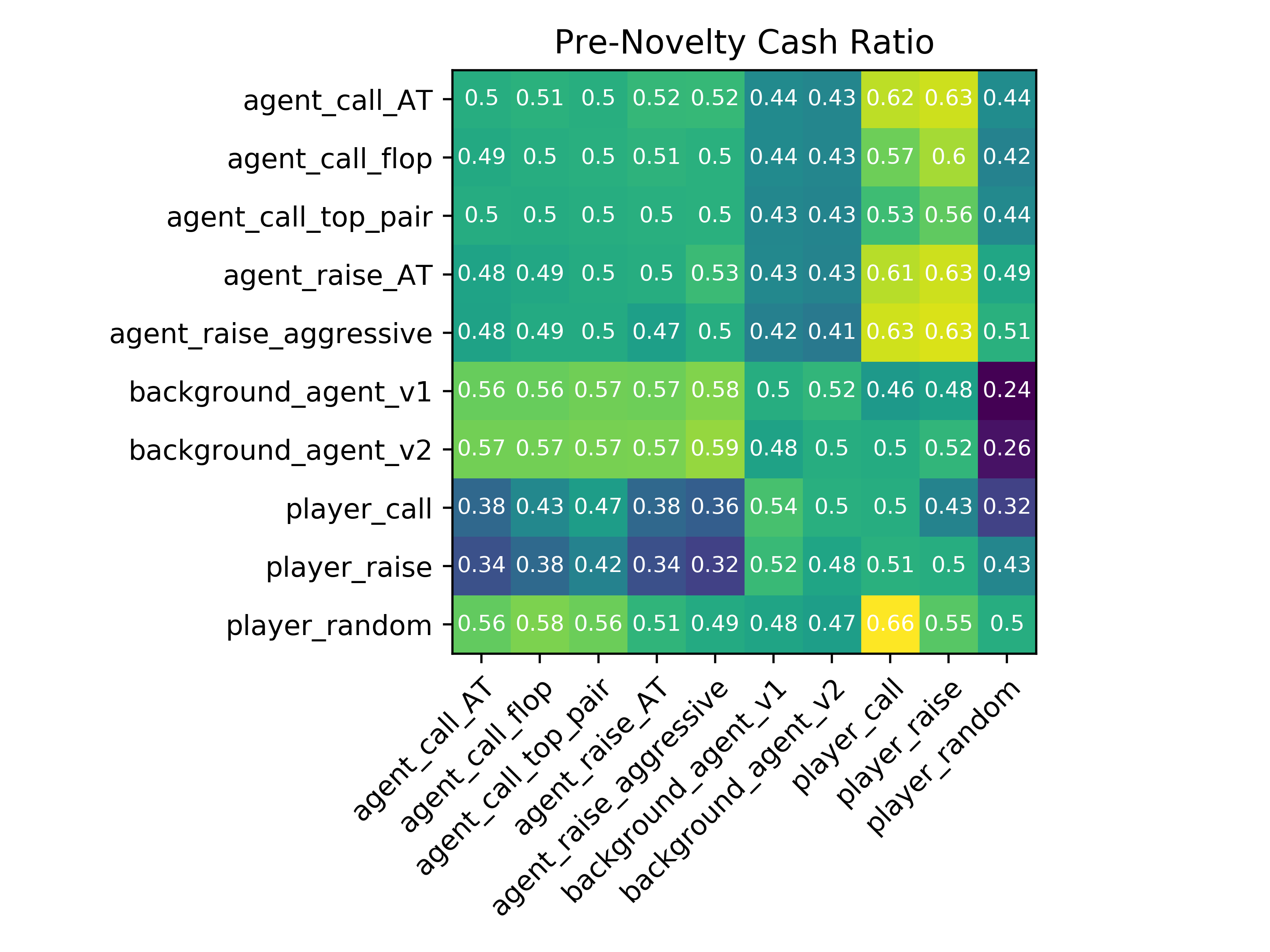}
\caption{Pre-novelty interaction matrix for Poker, showing the baseline cash ratio between agents.}
\label{fig:poker_pre}
\end{figure}

The pre-novelty interaction matrix in Figure \ref{fig:poker_pre} establishes the baseline performance of the ten Poker agents. We observe that agent efficacy is highly dependent on the specific opponent being faced. For instance, the background agent V1 generally maintains a cash ratio between 0.46 and 0.56 against most opponents, indicating relatively balanced matches; however, it exhibits a significant performance dip when playing against \textit{player\_random}, where the ratio drops to 0.24. A similar trend is evident for background agent V2, which remains within the 0.45 to 0.55 range across the majority of its match-ups but also struggles against \textit{player\_random}. In contrast, agents employing more extreme fixed strategies, such as \textit{player\_call} and \textit{player\_raise}, demonstrate high volatility, with cash ratios deviating substantially from the evenly matched baseline of 0.5 depending on the opponent's propensity to fold or remain in the pot.

\begin{figure}[h]
\centering
\includegraphics[scale=0.8]{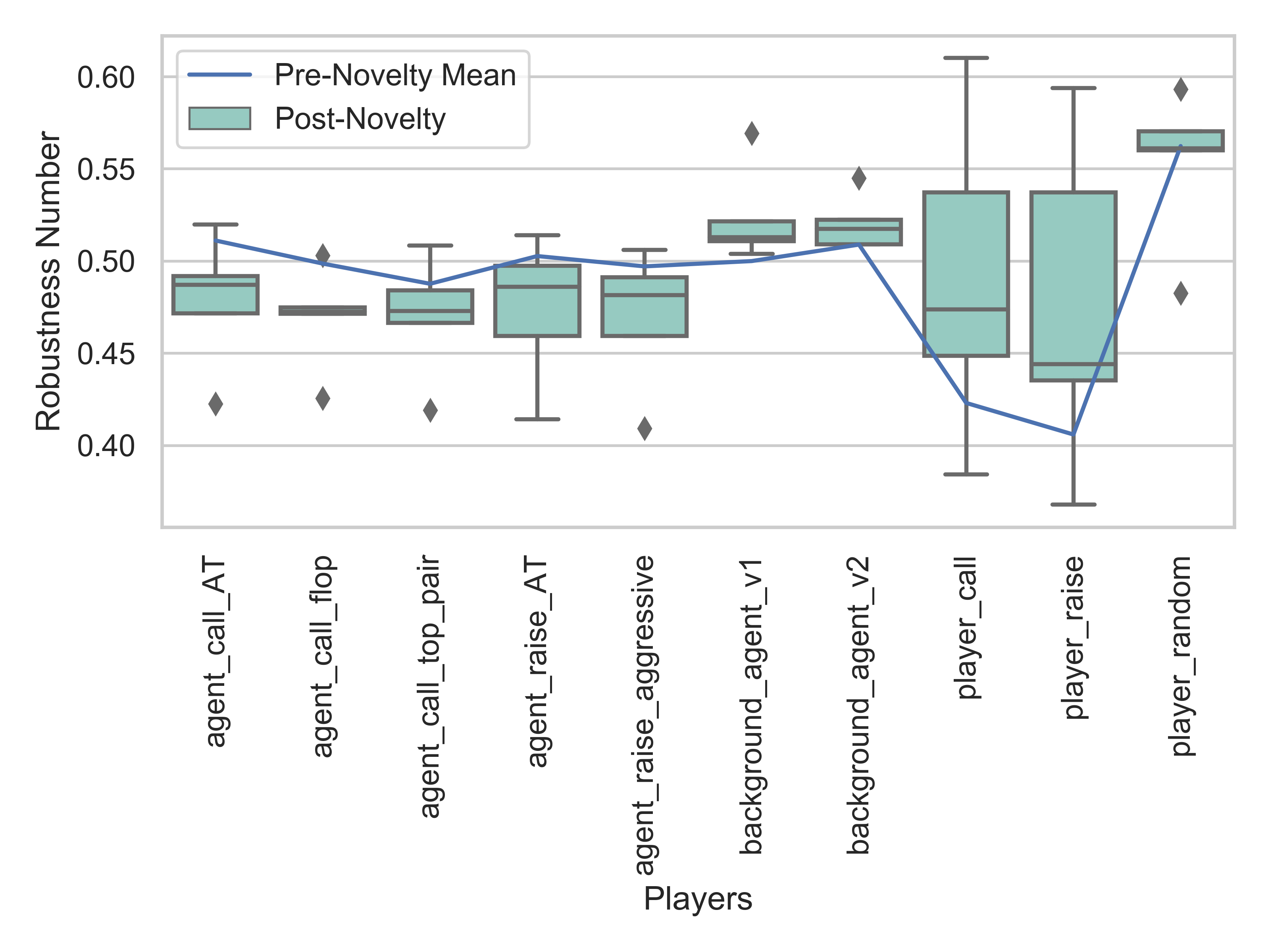}
\caption{Distribution of per-agent robustness for ten Poker agents. Each box plot summarizes performance across five distinct novelty scenarios.}
\label{fig:poker_agent}
\end{figure}

The distribution of per-agent robustness across the ten Poker agents is captured in Figure \ref{fig:poker_agent}. Similar to the patterns observed in the IPD domain, there is significant variation in how individual Poker strategies tolerate environmental novelty. Strategies such as \textit{player\_call} and \textit{player\_raise} exhibit high sensitivity to rule changes, with robustness values ranging from 0.45 to 0.55 and possessing large variance. This suggests that the performance of these simplistic betting strategies is highly contingent on the specific structural nature of the novelty introduced.

Conversely, more sophisticated or specialized agents, including \textit{raise\_AT} and \textit{raise\_aggressive}, maintain remarkably narrow robustness distributions, indicating a consistent performance profile across diverse novelty conditions. This trend is even more pronounced for the background agents and the random player, where the robustness values remain constrained to a very limited range. Notably, the random player achieves the highest average per-agent robustness. This finding, while potentially counterintuitive, reflects the inherent difficulty for rule-based or hand-ranking novelties to systematically degrade a strategy that lacks a deterministic or patterned response, thereby making it more invariant to specific environmental disruptions than more rigid agents.

\begin{figure}[h]
\centering
\includegraphics[scale=0.6]{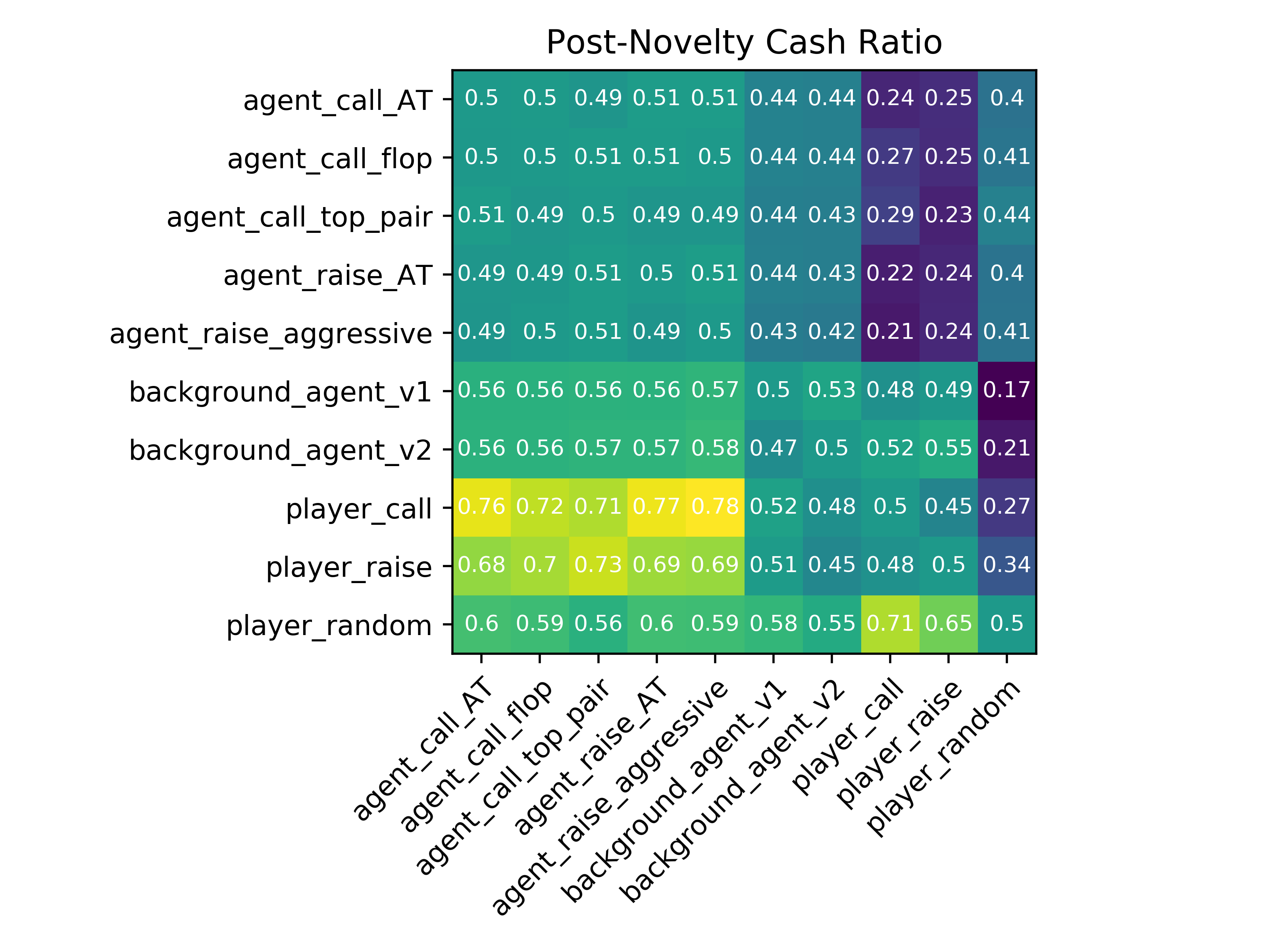}
\caption{Post-novelty cash ratio matrix under the "exchange hand" novelty scenario.}
\label{fig:poker_nov_exchange}
\end{figure}

\begin{figure}[h]
\centering
\includegraphics[scale=0.6]{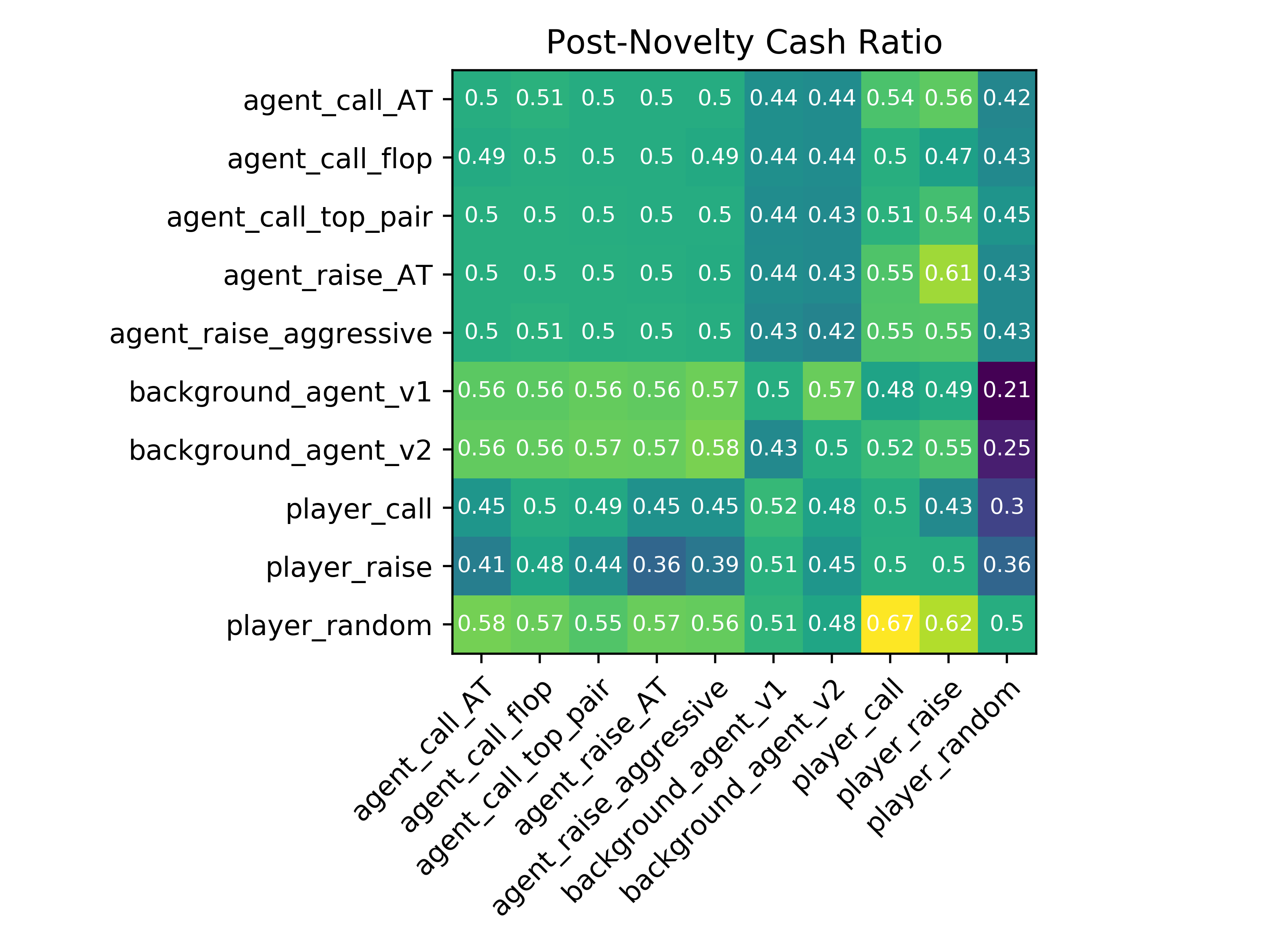}
\caption{Post-novelty cash ratio matrix under the "reorder hand rank" novelty scenario.}
\label{fig:poker_nov_reorder_hand}
\end{figure}

\begin{figure}[h]
\centering
\includegraphics[scale=0.6]{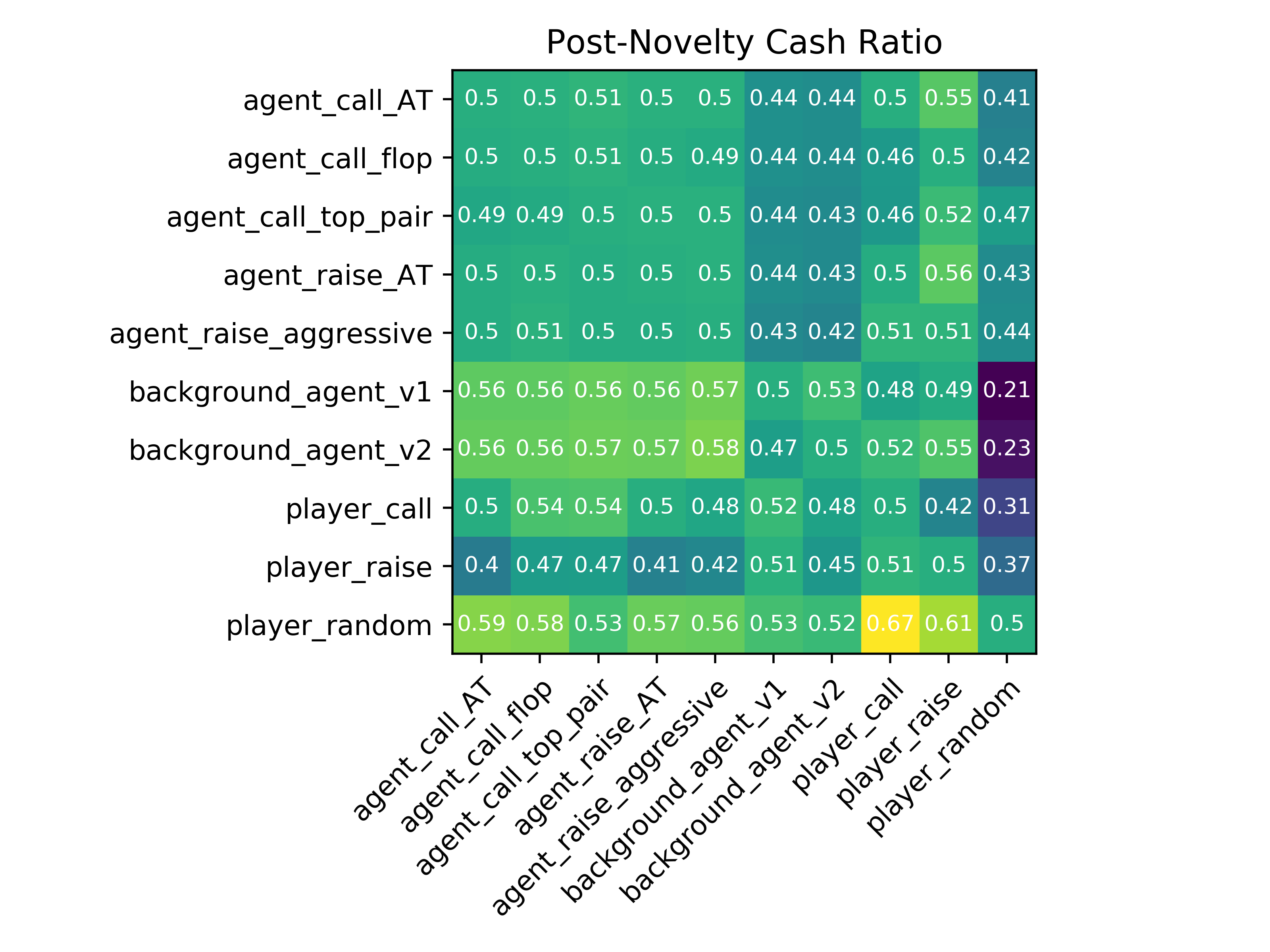}
\caption{Post-novelty cash ratio matrix under the "reorder number rank" novelty scenario.}
\label{fig:poker_nov_reorder_num}
\end{figure}

\begin{figure}[h]
\centering
\includegraphics[scale=0.6]{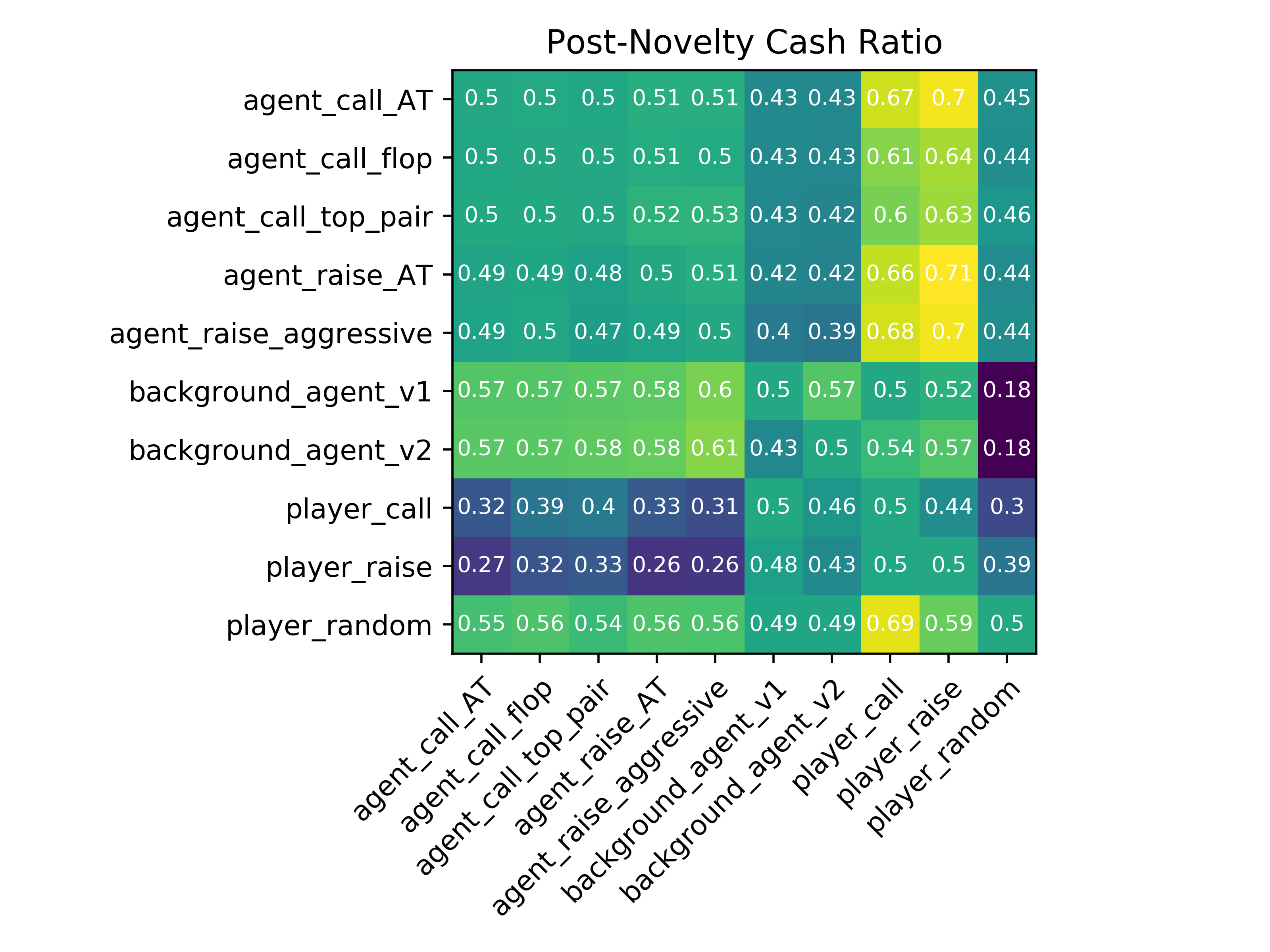}
\caption{Post-novelty cash ratio matrix under the "royal texas" novelty scenario.}
\label{fig:poker_nov_royal}
\end{figure}

\begin{figure}[h]
\centering
\includegraphics[scale=0.6]{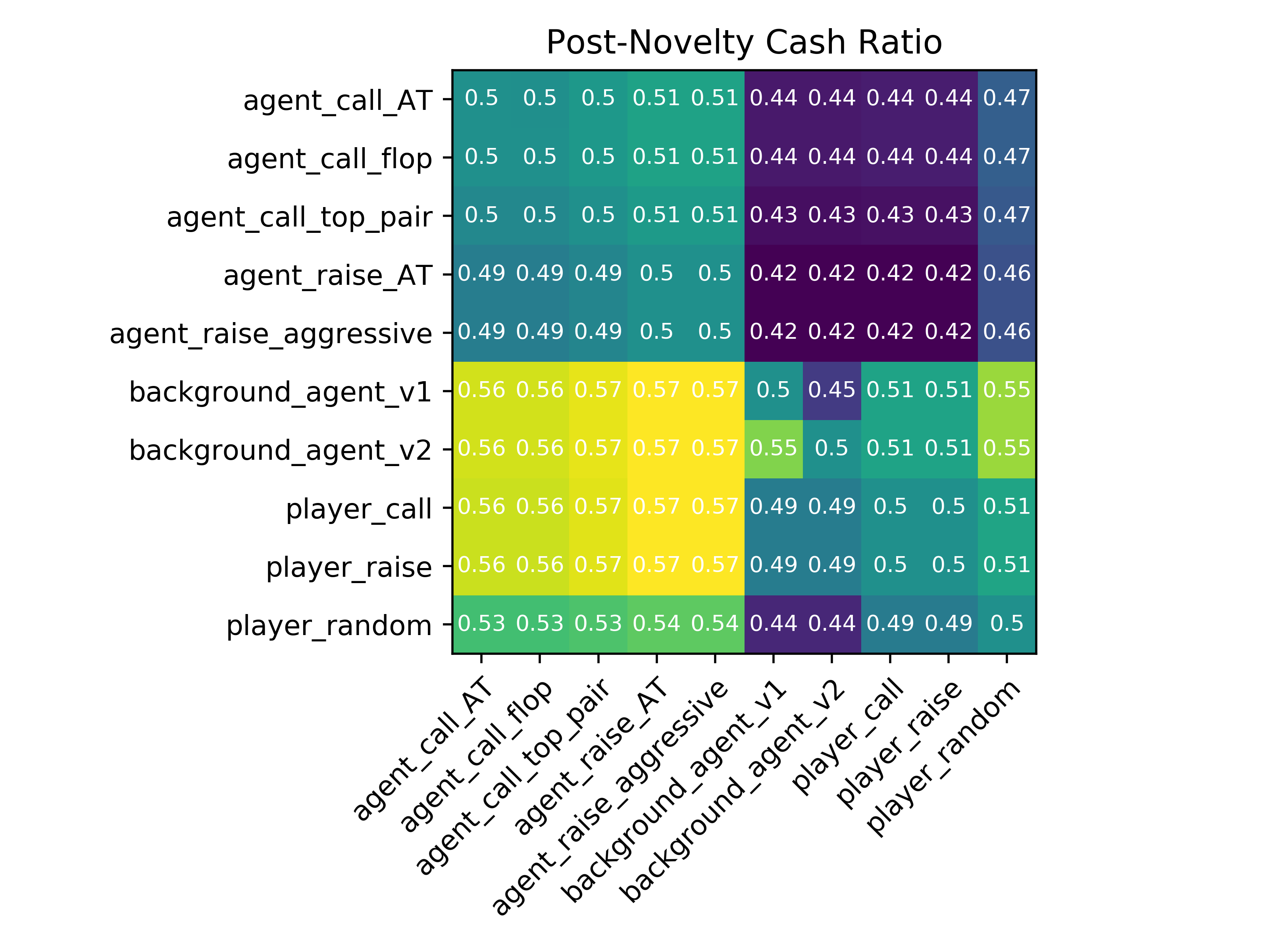}
\caption{Post-novelty cash ratio matrix under the "All-in/fold at flop round" novelty scenario.}
\label{fig:poker_nov_flop}
\end{figure}

The disparate impacts of specific novelty types are further evidenced in the post-novelty cash ratio matrices illustrated in Figures \ref{fig:poker_nov_exchange} through \ref{fig:poker_nov_flop}. Parallel to our findings in the IPD domain, the precise nature of the environmental perturbation dictates the resulting competitive landscape. For instance, the "exchange hand" novelty (Figure \ref{fig:poker_nov_exchange}) primarily disrupts the performance of simple strategies like \textit{player\_call} and \textit{player\_raise}, suggesting that hand-swapping mechanics specifically penalize rigid betting patterns. 

In contrast, the "reorder hand rank" (Figure \ref{fig:poker_nov_reorder_hand}) and "reorder number rank" (Figure \ref{fig:poker_nov_reorder_num}) scenarios exhibit a more muted impact relative to the baseline, though they still induce noticeable shifts for \textit{player\_raise}. The "Royal Texas" novelty (Figure \ref{fig:poker_nov_royal}) presents a markedly different matrix structure, as the truncation of the deck alters the probabilistic foundations of the game, thereby favoring agents whose strategies align with the higher hand densities. Collectively, these results emphasize that robustness is not merely a function of the agent's logic, but is deeply intertwined with the specific rules being modified. The interaction between agent architecture and novelty type determines the extent to which a competitive hierarchy is preserved or overturned.

\begin{figure}[h]
\centering
\includegraphics[scale=0.8]{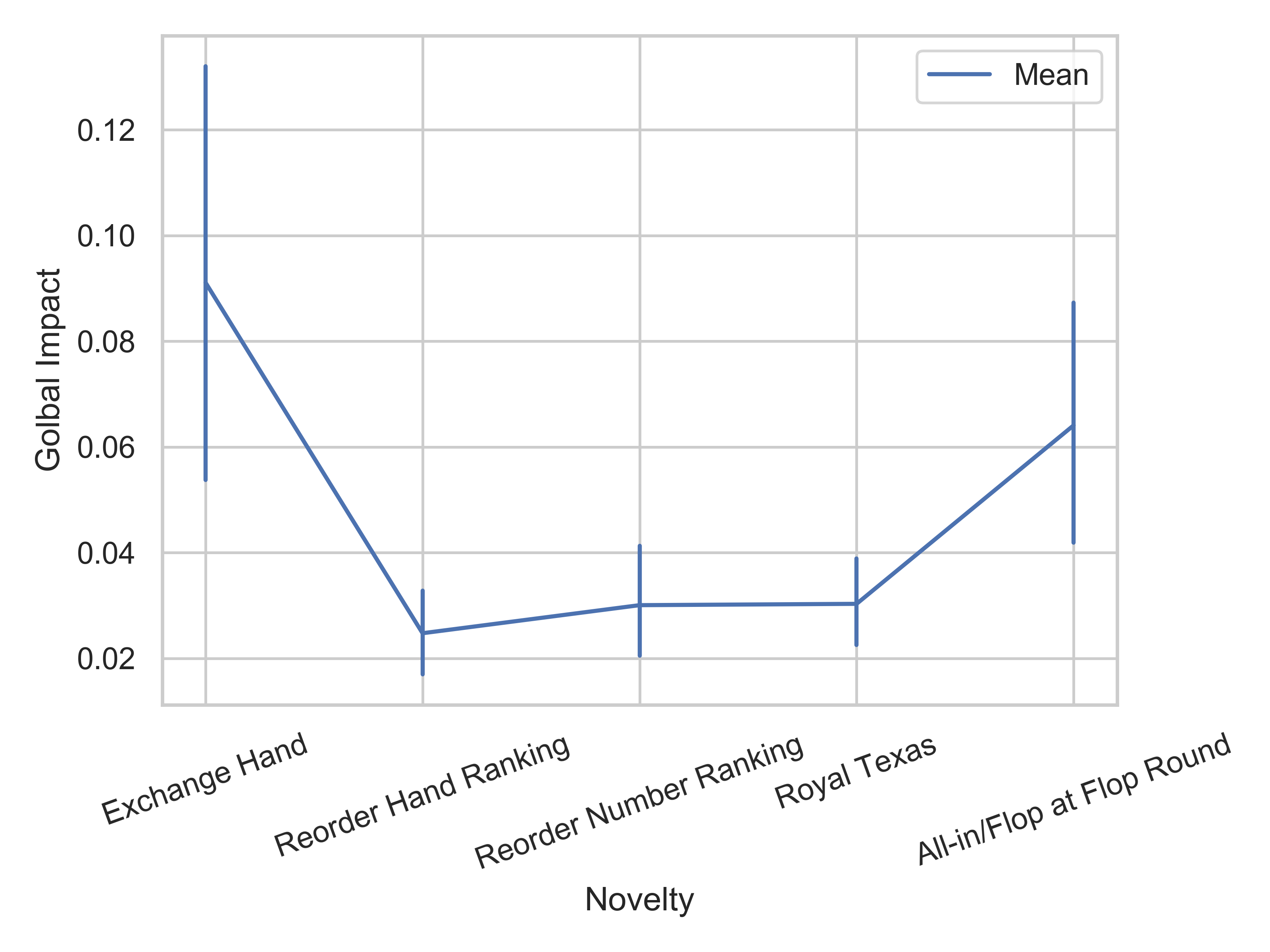}
\caption{The global impact of each novelty in the Poker domain (with standard errors).}
\label{fig:poker_impact}
\end{figure}

The global impact of the five Poker novelties is summarized in Figure \ref{fig:poker_impact}. In contrast to the IPD domain, where numerous novelties induced significant population-wide disruption, the Poker novelties generally exhibit relatively low global impact. The "exchange hand" and "All-in/fold at flop round" novelties are the primary exceptions, with global impact values of approximately 0.09 and 0.07, respectively. These scenarios also demonstrate higher variance, though the results remain statistically meaningful. The remaining three novelties—reorder hand rank, reorder number rank, and Royal Texas—show markedly lower impact, suggesting that the agent population is more resilient to these specific mechanical shifts. This difference between domains may be a consequence of the larger agent and novelty pools evaluated in the IPD experiments; it is plausible that a broader search space of Poker novelties would reveal a similar spectrum of high-impact disruptions. However, within the scope of these five injections, only the hand exchange and flop round constraints represent moderate to high disruptions to the competitive equilibrium.

\section{Discussion and Limitations}\label{sec:discussion}

The results presented in this work demonstrate that agent robustness is not a uniform trait but a highly context-dependent property that varies significantly across different classes of novelty. While our matrix-based methodology provides a rigorous framework for quantifying these effects, several limitations must be acknowledged. First, our study is confined to two-player zero-sum games. While IPD and Poker represent diverse information structures, it remains to be seen how these metrics scale to multi-agent cooperative or mixed-motive environments where the conservation of utility does not hold. In such settings, a novelty might not just shift performance but fundamentally alter the social welfare properties of the system.

Second, the novelties we injected were discrete and well-defined rule or payoff perturbations. In real-world open worlds, environmental changes may be continuous, gradual, or multi-faceted, making it difficult to pinpoint the exact onset or nature of the novelty. Furthermore, our analysis treats agents as black boxes with fixed strategies. This approach, while useful for establishing a baseline for the necessity of learning, does not account for the internal state representations or inductive biases that might make certain agents more naturally resilient than others. Our study is primarily descriptive rather than predictive; we can measure the impact of a novelty after the fact, but we lack a generative model to forecast how a previously unseen novelty will disrupt a specific strategic hierarchy without running the actual tournaments.

\section{Conclusion and Future Work}\label{sec:conclusion}

In this paper, we characterized the robustness of fixed-strategy agents in zero-sum open worlds using a domain-agnostic analytical framework. By evaluating a large corpus of agents across varied novelties in both IPD and Poker, we demonstrated that even highly optimized strategies can exhibit surprising fragility when their environmental assumptions are violated. These findings serve as a strong empirical justification for the development of advanced open-world learning systems that go beyond simple detection to active characterization and adaptation.

Future research should expand this investigation along several critical dimensions. A primary requirement is the inclusion of significantly more domains, especially those involving Large Language Models (LLMs) and the more recent wave of agentic AI systems. Assessing the robustness of models that rely on high-dimensional natural language representations or tool-use capabilities to environmental "hallucinations" or rule shifts—such as changes in API constraints or conversational protocols \cite{dialog1, evalGen}—is a vital next step. 

Furthermore, we aim to extend this work from a post-hoc analysis toward a predictive and prescriptive paradigm. This includes developing methods to automatically identify the robustness of an agent before deployment and to predict the global impact of a novelty based on its mathematical or semantic distance from the original environment. Ultimately, the most significant challenge remains the modification prong of open-world learning: creating algorithms that can not only detect and characterize novelty in stochastic environments but also autonomously update an agent's policy to mitigate performance degradation. Such advancements will be essential for deploying truly resilient autonomous systems in the unconstrained environments of the future.


\section{Authors' Contributions}

S.T. and H.L. conducted all experiments for this paper, implemented software, and synthesized many of the key results. M.K. wrote the original manuscript and supervised the effort. 

\section{Competing Interests}
There are no competing interests.

\section{Funding}

This work was funded by the Defense Advanced Research Projects Agency (DARPA) with award W911NF2020003 under the SAIL-ON program.



\begin{thebibliography}{99}

\bibitem{SAILON2021} Senator, Mr Ted. Science of artificial intelligence and learning for open-world novelty (sail-on). 2019.

\bibitem{ZS1} Pourpanah, Farhad, Abdar, Moloud, Luo, Yuxuan, Zhou, Xinlei, Wang, Ran, Lim, Chee Peng, Wang, Xi-Zhao, Wu, QM Jonathan. A review of generalized zero-shot learning methods. IEEE transactions on pattern analysis and machine intelligence. 2022.

\bibitem{ZS2} Xian, Yongqin, Lampert, Christoph H, Schiele, Bernt, Akata, Zeynep. Zero-shot learning—a comprehensive evaluation of the good, the bad and the ugly. IEEE transactions on pattern analysis and machine intelligence, 41(9), pp. 2251--2265. 2018.

\bibitem{axelrod1980effective} Axelrod, Robert. Effective choice in the prisoner's dilemma. Journal of conflict resolution, 24(1), pp. 3--25. 1980.

\bibitem{boult2019learning} Boult, Terrance E, Cruz, Steve, Dhamija, Akshay Raj, Gunther, Manuel, Henrydoss, James, Scheirer, Walter J. Learning and the unknown: Surveying steps toward open world recognition. In: Proceedings of the AAAI conference on artificial intelligence, pp. 9801--9807. 2019.

\bibitem{dialog1} Fuad, Ahlam, Al-Yahya, Maha. Recent developments in arabic conversational AI: A literature review. IEEE Access. 2022.

\bibitem{eval4} Gamage, Chathura, Pinto, Vimukthini, Xue, Cheng, Doctor, K, Aha, David. Novelty generation framework for AI agents in angry birds style physics games. In: 2021 IEEE Conference on Games (CoG), pp. 1--8. 2021.

\bibitem{eval5} Horner, K. Polycraft team to lay groundwork for smarter AI. 2020.

\bibitem{eval6} Jiang, Xia, Cooper, Gregory F, Neill, Daniel B. Generalized AMOC curves for evaluation and improvement of event surveillance. In: AMIA Annual Symposium Proceedings, pp. 281. 2009.

\bibitem{eval7} Kejriwal, Mayank, Thomas, Shilpa. A multi-agent simulator for generating novelty in monopoly. Simulation Modelling Practice and Theory, 112, pp. 102364. 2021.

\bibitem{eval8} Molineaux, Matthew, Dannenhauer, Dustin. An Environment Transformation-based Framework for Comparison of Open-World Learning Agents.

\bibitem{eval9} Pinto, Vimukthini, Renz, Jochen, Xue, Cheng, Doctor, K, Aha, D. Measuring the Performance of Open-World AI Systems. 2020.

\bibitem{evalGen} Chen, Mark, Tworek, Jerry, Jun, Heewoo, Leike, Jan, Radford, Alec, Wang, Jiayi, Mishchenko, Kai, Liu, Juhao, Plappert, Matthias, Chou, Jason, Mitter, Sascha, Kim, Chris, Hilton, Jacob, Nakano, Reiichiro, Heisenreich, Christopher. Evaluating large language models trained on code. arXiv preprint arXiv:2107.03374. 2021.

\bibitem{goss2023polycraft} Goss, Stephen A, Steininger, Robert J, Narayanan, Dhruv, Alicea, Brian, Dickens, John, Roberts, Isaac, Banerjee, Devendra, Dey, Amrita, Sinha, Abhijit, Li, Hongyu, Kejriwal, Mayank, Riedl, Mark O. Polycraft World AI Lab (PAL): An Extensible Platform for Evaluating Artificial Intelligence Agents. arXiv preprint arXiv:2301.11891. 2023.

\bibitem{killick2014changepoint} Killick, Rebecca, Eckley, Idris. changepoint: An R package for changepoint analysis. Journal of statistical software, 58(3), pp. 1--19. 2014.

\bibitem{li2011engineering} Li, Jiawei, Hingston, Philip, Kendall, Graham. Engineering design of strategies for winning iterated prisoner's dilemma competitions. IEEE Transactions on Computational Intelligence and AI in Games, 3(4), pp. 348--360. 2011.

\bibitem{main1} Heaven, Douglas, others. Why deep-learning AIs are so easy to fool. Nature, 574(7777), pp. 163--166. 2019.

\bibitem{main2} Marcus, Gary. Deep learning: A critical appraisal. arXiv preprint arXiv:1801.00631. 2018.

\bibitem{main4} Bulusu, Saikiran, Kailkhura, Bhavya, Li, Bo, Varshney, Pramod K, Song, Dawn. Anomalous example detection in deep learning: A survey. IEEE Access, 8, pp. 132330--132347. 2020.

\bibitem{main5} Kejriwal, Mayank, Kildebeck, Eric, Shrivastava, Abhinav. Designing Artificial Intelligence for Open Worlds. 2022.

\bibitem{main8} Musliner, David J, Pelican, Michael JS, McLure, Matthew, Johnston, Steven, Freedman, Richard G, Knutson, Corey. OpenMIND: Planning and adapting in domains with novelty. In: Proc. Ninth Conf. on Advances in Cognitive Systems. 2021.

\bibitem{main9} Kejriwal, Mayank, Thomas, Shilpa. A multi-agent simulator for generating novelty in monopoly. Simulation Modelling Practice and Theory, 112, pp. 102364. 2021.

\bibitem{main11} Cincotti, Alessandro, Iida, Hiroyuki, Yoshimura, Jin. Refinement and complexity in the evolution of chess. In: Information Sciences 2007, pp. 650--654. World Scientific, 2007.

\bibitem{main12} Berger, Emily Rita, Dubbs, Alexander. Winning Strategies in Multimove Chess (i, j). Journal of Information Processing, 23(3), pp. 272--275. 2015.

\bibitem{main13} Silver, David, Hubert, Thomas, Schrittwieser, Julian, Antonoglou, Ioannis, Lai, Matthew, Guez, Arthur, Lanctot, Marc, Sifre, Laurent, Kumaran, Dharshan, Graepel, Thore, others. A general reinforcement learning algorithm that masters chess, shogi, and Go through self-play. Science, 362(6419), pp. 1140--1144. 2018.

\bibitem{main14} Tu, James, Li, Huichen, Yan, Xinchen, Ren, Mengye, Chen, Yun, Liang, Ming, Bitar, Eilyan, Yumer, Ersin, Urtasun, Raquel. Exploring adversarial robustness of multi-sensor perception systems in self driving. In: Conference on Robot Learning, pp. 1013--1024. 2021.

\bibitem{main15} Naudé, Wim. Artificial intelligence vs COVID-19: limitations, constraints and pitfalls. AI \& society, 35, pp. 761--765. 2020.

\bibitem{main20} Wang, Yaqing, Yao, Quanming, Kwok, James T, Ni, Lionel M. Generalizing from a few examples: A survey on few-shot learning. ACM computing surveys (CSUR), 53(3), pp. 1--34. 2020.

\bibitem{mittaloptimal} Mittal, Shashi, Deb, Kalyanmoy. Optimal Strategies of the Iterated Prisoner's Dilemma Problem for Multiple Conflicting Objectives. Indian Institute of Technology, Kanpur, India.

\bibitem{opp7} Langley, Pat. Constraints on Theories of Open-World Learning. 2022.

\bibitem{opp8} Boult, Terrance, Grabowicz, Przemyslaw, Prijatelj, Derek, Stern, Roni, Holder, Lawrence, Alspector, Joshua, Jafarzadeh, Mohsen M, Ahmad, Toqueer, Dhamija, Akshay, Li, Chunchun, others. Towards a unifying framework for formal theories of novelty. In: Proceedings of the AAAI Conference on Artificial Intelligence, 35(17), pp. 15047--15052. 2021.

\bibitem{opp11} Lakkaraju, Himabindu, Kamar, Ece, Caruana, Rich, Horvitz, Eric. Identifying unknown unknowns in the open world: Representations and policies for guided exploration. In: Proceedings of the AAAI Conference on Artificial Intelligence, 31(1). 2017.

\bibitem{press2012iterated} Press, William H, Dyson, Freeman J. Iterated Prisoner's Dilemma contains strategies that dominate any evolutionary opponent. Proceedings of the National Academy of Sciences, 109(26), pp. 10409--10413. 2012.

\bibitem{zhang2018overview} Zhang, Yu, Yang, Qiang. An overview of multi-task learning. National Science Review, 5(1), pp. 30--43. 2018.

\end{thebibliography}

%
%
%
%
%

\newpage

\section{Appendix A: Prisoner's Dilemma Agents Used in Experimental Study}

This appendix provides a detailed overview of the thirty agents utilized in our Iterated Prisoner's Dilemma (IPD) experiments. The agents range from simple fixed-strategy baselines, such as Cooperator and Defector, to more complex adaptive algorithms and reinforcement learning models. Table \ref{table:ipd_agents_1} summarizes the first fifteen agents, including fundamental strategies like Tit for Tat and various copier or punisher archetypes. 

Table \ref{table:ipd_agents_2} details the remaining fifteen agents, which include adaptive variants and multiple Q-learning implementations. By testing this diverse population against a wide array of payoff novelties, we are able to characterize how different levels of strategic complexity respond to environmental perturbations.

\begin{table}[p]
\centering
\small
\begin{tabular}{|l|l|p{9.5cm}|}
\hline
\textbf{ID} & \textbf{Agent} & \textbf{Strategy Description} \\ \hline
 1 & Cooperator & An agent that only cooperates. \\ \hline
 2 & Defector & An agent that only defects. \\ \hline
 3 & Tit for Tat & Imitates the previous action of the opponent. \\ \hline
 4 & Alternator & Alternates between cooperating and defecting. \\ \hline
 5 & Adaptive & Uses a specific sequence and then plays the best strategy. \\ \hline
 6 & Grudger & Cooperates initially but defects if the opponent ever defects. \\ \hline
 7 & Average Copier & Cooperates based on historical cooperation ratio. \\ \hline
 8 & Appeaser & Switches every time the opponent defects. \\ \hline
 9 & Firm but Fair & Cooperates but defects if payoffs are low. \\ \hline
 10 & First by Anonymous & Uniformly random cooperation (30-70\%). \\ \hline
 11 & Tit for 2 Tats & Defects only after two opponent defections. \\ \hline
 12 & 2 Tits for Tat & Defects twice for each opponent defection. \\ \hline
 13 & Defector Hunter & Switches if it detects persistent defection. \\ \hline
 14 & Punisher & Proportional punishment rounds for defection. \\ \hline
 15 & Inverse Punisher & Penalizes the opponent for cooperating. \\ \hline
\end{tabular}
\caption{Summary of IPD Agents (1-15)}
\label{table:ipd_agents_1}
\end{table}

\begin{table}[p]
\centering
\small
\begin{tabular}{|l|l|p{9.5cm}|}
\hline
\textbf{ID} & \textbf{Agent} & \textbf{Strategy Description} \\ \hline
 16 & Adaptor Brief & Adaptive agent; short interactions. \\ \hline
 17 & Adaptor Long & Adaptive agent; long interactions. \\ \hline
 18 & Adaptive Tit for Tat & Tit-for-tat with an adaptive rate. \\ \hline
 19 & Anti Tit for Tat & Plays the opposite of opponent's move. \\ \hline
 20 & Bully & Starts with defection; plays opposite of opponent. \\ \hline
 21 & Gradual & Incremental punishment for defection. \\ \hline
 22 & Gradual Killer & Fixed sequence and conditional behavior. \\ \hline
 23 & Easy Go & Cooperates if the opponent ever defects. \\ \hline
 24 & Handshake & Handshake sequence to identify similar agents. \\ \hline
 25 & Hard Prober & Probe sequence followed by conditional behavior. \\ \hline
 26 & Arrogant Q Learner & Aggressive Q-learner. \\ \hline
 27 & Cautious Q Learner & Deliberate Q-learner with more look-ahead. \\ \hline
 28 & Hesistant Q Learner & Moderate Q-learner. \\ \hline
 29 & Resurrection & Conditional defection based on recent rounds. \\ \hline
 30 & Limited Retaliate & Retaliates until a win-loss limit is hit. \\ \hline
\end{tabular}
\caption{Summary of IPD Agents (16-30)}
\label{table:ipd_agents_2}
\end{table}

\clearpage
\section{Appendix B: Poker Strategies and Scenario Details}

This appendix contains the specifics of the agents and the novelty injections used in the Poker domain. Our Poker experiments involve ten distinct strategies, ranging from simple rule-based agents to more sophisticated background models. The behavior of these agents is defined across different stages of the game. Table \ref{table:poker_pre_flop} and Table \ref{table:poker_flop} describe the pre-flop and flop round strategies, respectively, for all ten agents. Table \ref{table:poker_turn_river} provides additional details for the background agents during the turn and river rounds.

In addition to the agent strategies, we also describe the specific novelties injected into the Poker environment to test agent robustness. Table \ref{table:poker_novelties} summarizes these five novelties, which include structural changes to the deck, hand rankings, and action constraints. These perturbations allow us to observe how strategies optimized for standard heads-up Texas Hold'em perform when fundamental game rules are altered.

\begin{table}[h]
\centering
\small
\begin{tabular}{|l|l|p{9cm}|}
\hline
\textbf{ID} & \textbf{Agent} & \textbf{Pre-Flop Round Description} \\ \hline
 1 & Raise & Raises if cash > 2x previous bet. \\ \hline
 2 & Call & Calls if cash $\ge$ previous bet. \\ \hline
 3 & Random & Random allowable actions (fold, call, raise, all-in). \\ \hline
 4 & Call-AT & Calls for pairs 7-A or 10-A combinations. \\ \hline
 5 & Raise-AT & Raises for pairs 7-A or 10-A combinations. \\ \hline
 6 & Raise-Aggressive & Raises 10x for pairs 7-A or 10-A combinations. \\ \hline
 7 & Call-TopPair & Calls only for cards between 11 (Jack) and Ace. \\ \hline
 8 & Call-Flop & Calls for pairs 7-A or 10-A combinations. \\ \hline
 9 & Background V1 & Uses categorical risk management and percent limits. \\ \hline
 10 & Background V2 & Same strategy as Background V1. \\ \hline
\end{tabular}
\caption{Poker Pre-Flop Strategies}
\label{table:poker_pre_flop}
\end{table}

\begin{table}[h]
\centering
\small
\begin{tabular}{|l|l|p{9cm}|}
\hline
\textbf{ID} & \textbf{Agent} & \textbf{Flop Round Description} \\ \hline
 1 & Raise & Doubles bet; folds if opponent all-in. \\ \hline
 2 & Call & Calls bet; folds if opponent all-in. \\ \hline
 3 & Random & Random allowable actions. \\ \hline
 4 & Call-AT & Calls bet; folds if opponent all-in. \\ \hline
 5 & Raise-AT & Doubles bet; folds if opponent all-in. \\ \hline
 6 & Raise-Aggressive & Raises 10x; folds if opponent all-in. \\ \hline
 7 & Call-TopPair & Calls bet; folds if opponent all-in. \\ \hline
 8 & Call-Flop & Calls if $\ge 3$ cards have same values or suits. \\ \hline
 9 & Background V1 & All-in if in top 1-3 pairs; strategic folding thresholds. \\ \hline
 10 & Background V2 & Same strategy as Background V1. \\ \hline
\end{tabular}
\caption{Poker Flop Round Strategies}
\label{table:poker_flop}
\end{table}

\begin{table}[h]
\centering
\small
\begin{tabular}{|l|l|p{9cm}|}
\hline
\textbf{ID} & \textbf{Agent} & \textbf{Turn/River Round Description} \\ \hline
 9 (Turn) & Background V1 & Checks if possible, otherwise folds. \\ \hline
 10 (Turn) & Background V2 & All-in for strong suits/values; or calls up to 2x min bet. \\ \hline
 9 (River) & Background V1 & Same strategy as turn round. \\ \hline
 10 (River) & Background V2 & All-in for Royal Flush, etc.; calls up to 2x min bet. \\ \hline
\end{tabular}
\caption{Poker Turn and River Strategies (Background Agents)}
\label{table:poker_turn_river}
\end{table}

\begin{table}[h]
\centering
\small
\begin{tabular}{|l|l|p{9.5cm}|}
\hline
\textbf{ID} & \textbf{Novelty} & \textbf{Description} \\ \hline
 1 & Exchange Hand & Hands are moved between active players before showdown. \\ \hline
 2 & Hand Ranking & Shuffles hand hierarchy (e.g., High Card > Full House). \\ \hline
 3 & Number Ranking & Shuffles card values (e.g., 5 > Ace). \\ \hline
 4 & Royal Texas & Deck is limited to cards with values 10 and above. \\ \hline
 5 & Action Constraints & Call/Fold (pre-flop) or All-in/Fold (flop) restrictions. \\ \hline
\end{tabular}
\caption{Summary of Poker Novelties}
\label{table:poker_novelties}
\end{table}



\end{document}